\documentclass{svproc}
\usepackage{url}

\usepackage{graphicx}
\graphicspath{{figures/}}

\usepackage{xcolor}

\begin{document}
\mainmatter              % start of a contribution
\title{AI for Agile development: a Meta-Analysis}
\titlerunning{AI-assisted Agile}  % abbreviated title (for running head)

\author{Beatriz Cabrero-Daniel}
\institute{University of Gothenburg, Gothenburg 41756, Sweden,\\
\email{beatriz.cabrero-daniel@gu.se}}

% \author{Beatriz Cabrero-Daniel}
% \affiliation{%
%   \institution{University of Gothenburg}
%   \streetaddress{Hörselgången 5}
%   \city{Gothenburg}
%   \country{Sweden}
%   \postcode{41756}
% }

\maketitle              % typeset the title of the contribution

\begin{abstract}
% CONTEXT
This study explores the benefits and challenges of integrating Artificial Intelligence with Agile software development methodologies, focusing on improving continuous integration and delivery. 
% METHODS
A systematic literature review and longitudinal meta-analysis of the retrieved studies was conducted to analyse the role of Artificial Intelligence and it's future applications within Agile software development. The review helped identify critical challenges, such as the need for specialised socio-technical expertise. 
% RESULTS
While Artificial Intelligence holds promise for improved software development practices, further research is needed to better understand its impact on  processes and practitioners, and to address the indirect challenges associated with its implementation.
\keywords{agile software development, artificial intelligence, meta-analysis}
\end{abstract}

%%%%%%%%%%%%%%%%%%%%%%%%%%%%%%%%%%%%%%%%%%%

\section{Introduction}

% THERE IS A TREND TO INTEGRATE AI BECAUSE IT HELPS SOFTWARE DEVELOPMENT
Certain Agile software development processes, such as task allocation, backlog prioritisation, test case reduction and prioritisation, are still mostly manual processes~\cite{singh2020technique}.The ever-changing market trends has made it necessary to release software faster while ensuring its quality, and AI can effectively be (and is increasingly being) integrated with Agile software development methodologies to enhance some of these processes: from test generators to user story conflict detectors~\cite{khanna2021artificial,lariosvargas2022}, the use of AI in Agile software development methodologies has gained popularity in recent years, particularly in the automotive sector where safety-critical software applications are developed~\cite{Boualouache2023}.

% AI CAN IMPROVE SOFTWARE DEVELOPMENT
This trend is reflected in the literature. For instance, inexperienced staff is one of the critical challenges of software development~\cite{khan2016systematic} but recent advances in the field of AI might help mitigate these human-factors-related challenges: novice software architects could collaborate with generic Large Language Models (LLMs) tools, such as ChatGPT\footnote{Visit \url{https://chat.openai.com/chat}}, in tasks such as architectural analysis or evaluation of services~\cite{ahmad2023towards}. In the future, LLM tools could even lead the software engineering tasks with appropriate human-oversight mechanisms~\cite{ahmad2023towards,AIact,yasolga2023}.

% GOAL OF PAPER
This paper reports the trends shown in the literature regarding the challenges of Agile software development \textbf{(RQ1)} and the usage of AI to mitigate them \textbf{(RQ2)}.
% \begin{itemize}
%     \item \textbf{RQ1)} What are the current challenges of Agile software development?
%     \item \textbf{RQ2)} What are the common uses of AI within Agile software development?
%     \item \textbf{RQ3)} What are the challenges for AI-assisted Agile software development and mitigation strategies in the literature? 
% \end{itemize}
This meta-analysis draws the bigger picture of the field of AI for Agile software development and reflects on potential future challenges directly or indirectly related to the integration of AI into Agile practices \textbf{(RQ3)}.

\section{Background}

% Benefits of implementing AI in Agile software development.
Lean software development leads to both decreased lead and development times and increased productivity while increasing customer satisfaction on the products~\cite{rodriguez2019advances}. In turn, the integration of AI in Agile can improve the performance of Agile practices, e.g., assisting in risk management~\cite{khanna2021artificial}, task allocation~\cite{william2021task}, supporting security practices~\cite{lariosvargas2022}, code analysis~\cite{machalica2019predictive}, and test case prioritisation in continuous integration settings~\cite{da2022machine,marijan2020neural,kumar2022review,william2021task,singh2021framework}. AI ca also benefit Agile team work by tracking team performance~\cite{ameta2022scaled} and mitigate potential human-errors with software defect detection and automatic correction~\cite{Boualouache2023}.

% Challenges of implementing AI in Agile software development.
Nevertheless, the related literature highlights several significant challenges related to the integration of AI with Agile software development methodologies~\cite{lariosvargas2022}. Challenges are not limited to the software development process, but also include AI model management, deployment, and monitoring \cite{granlund2021mlops,rahman2022challenges,makinen2021needs}. Machine Learning (ML) misbehaviour detectors are therefore critical~\cite{Boualouache2023} and so is implementing secure software development pipelines~\cite{lariosvargas2022} and risk management frameworks~\cite{khanna2021artificial}. AI also poses indirect challenges that can affect safety-critical software products and must therefore be carefully considered in order to successfully integrate AI into Agile software development~\cite{Boualouache2023}, e.g., the appropriateness and treatment of data sources, specially in the case of big data~\cite{Pulikottil2023}.

% Human-related challenges
Context- and human-factors must also be considered. For instance, for backlog prioritisation, AI models need to be aligned with project goals~\cite{kumar2022review}; this is difficult for practitioners since projects can be from many different industries, from sports to financial~\cite{pang2021promotion,gan2021automated}. Many other challenges have been reported in the literature, e.g., lack of standardised processes, socio-technical limitations, or the scarcity of human expertise~\cite{ahmad2023towards}. AI engineers must therefore have a clear understanding of the software development life cycle and agile practices~\cite{meesters2022ai,mukhtar2013hybrid}.

\section{Methodology} \label{sec:method}

% \textbf{Selection of primary studies.} 
A review protocol, described in this section, was defined to search the outcomes of previous research. Prior to the review, preliminary searches aimed at both identifying existing systematic reviews and meta-analysis, as well as assessing the volume of potentially relevant studies. Based on these efforts, queries for manual and keyword automated searches were crafted. For the selection of primary studies, we queried four formal databases (i.e., ACM Digital Library, Springer, IEEE Xplore, and ScienceDirect\footnote{Multiple queries had to be used to account for the limit of 8 Boolean operators.}) and one engine (i.e., Google Scholar) that indexes pre-print servers (e.g., arXiv) to look for papers about \textit{(Artificial Intelligence OR related topics) AND (Agile software development OR Continuous Integration-related topics)}. A total of 5446 (101 from IEEE, 1106 from ACM DL, 2747 from ScienceDirect, and 1492 from Springer) papers were identified through database searching and retrieved, partly thanks to an ad hoc web crawler based on AHK\footnote{Visit \url{https://pypi.org/project/ahk/}}. 31 records from Google Scholar were added (no duplicates).

% Full-text articles assessed for eligibility & n \\
% Studies included in qualitative synthesis & n \\
% Studies included in quantitative synthesis (meta-analysis) & n \\

% \textbf{Study selection and data extraction.} 
Given the need to study temporary trends in the literature in a longitudinal meta-analysis, studies were not excludes based on the publication date. Regardless of the publication date, all selected papers (i) are peer-reviewed articles (6+ pages) in English with full-text available, report either (ii) AI-based mitigation strategies for challenges in Agile software development or (iii) propose AI tools for improving the efficiency of Agile practices. Therefore, studies about Agile methodologies for AI-related software development were excluded. %, and (iv) use empirical research methods to verify the proposed strategies and tools
% \textbf{Data synthesis.} 
Records were first screened based on title, and on abstract in a second phase. After screening, 377 papers had been selected from which data was extracted (see \textbf{Appendix A}). The themes arising from this subset of papers were used to structure the presentation of the results. Statistical techniques, described in Section~\ref{sec:results}, were then use to quantitatively synthesise the results of our search.

\section{Categories of Agile development challenges (RQ1)} \label{sec:results}

To address \textbf{RQ1}, we analysed the current trends in the literature regarding the challenges in Agile software development and divided them into seven groups. On the one hand, challenges related to (i) user stories and use cases include the unavailability of user stories and the difficulty of assessing their quality. To address these challenges, value-driven software processes can guide decision making~\cite{rodriguez2019advances}. Therefore, it is important to involve providers and customers early in the co-creation processes~\cite{SJODIN2020478}. Feature co-validation will also determine what can be reused and what requires customised development~\cite{SJODIN2020478}. %reusingcode,client

Other challenges involve (ii) estimating the time and effort to complete a task or a project, and (iii) scaling Agile. Technical challenges (iv) encompass several areas such as developing secure software systems, handling requirements dependencies, limited research focusing on non-functional requirements, lack of automation approach, regression testing, and uncertainty in understanding how the system should interact with the user to achieve the intended benefit. Project artefacts (e.g., epics, tasks), revision history, workflow data logs, and other backlogs, are often managed through project management tools~\cite{Katarzyna2021}.

On the other hand, (v) team-related challenges include unclear responsibilities, managing dependencies, interacting asynchronously, identifying needs that lead to stakeholder satisfaction, dealing with inexperienced practitioners, and involving customers in Agile processes. Some of these challenges can result in more unplanned tasks and sub-tasks in the backlog, longer running stories, irregular burn-down charts, and higher deviation from estimated time. Similarly, challenges related to (vi) value creation and prioritisation involve defining and ranking non-functional needs, prioritising requirements that align with commercial goals, dealing with ambiguous requirements, and focusing on value creation. 

\begin{figure}
    \centering
    \includegraphics[width=.8\linewidth]{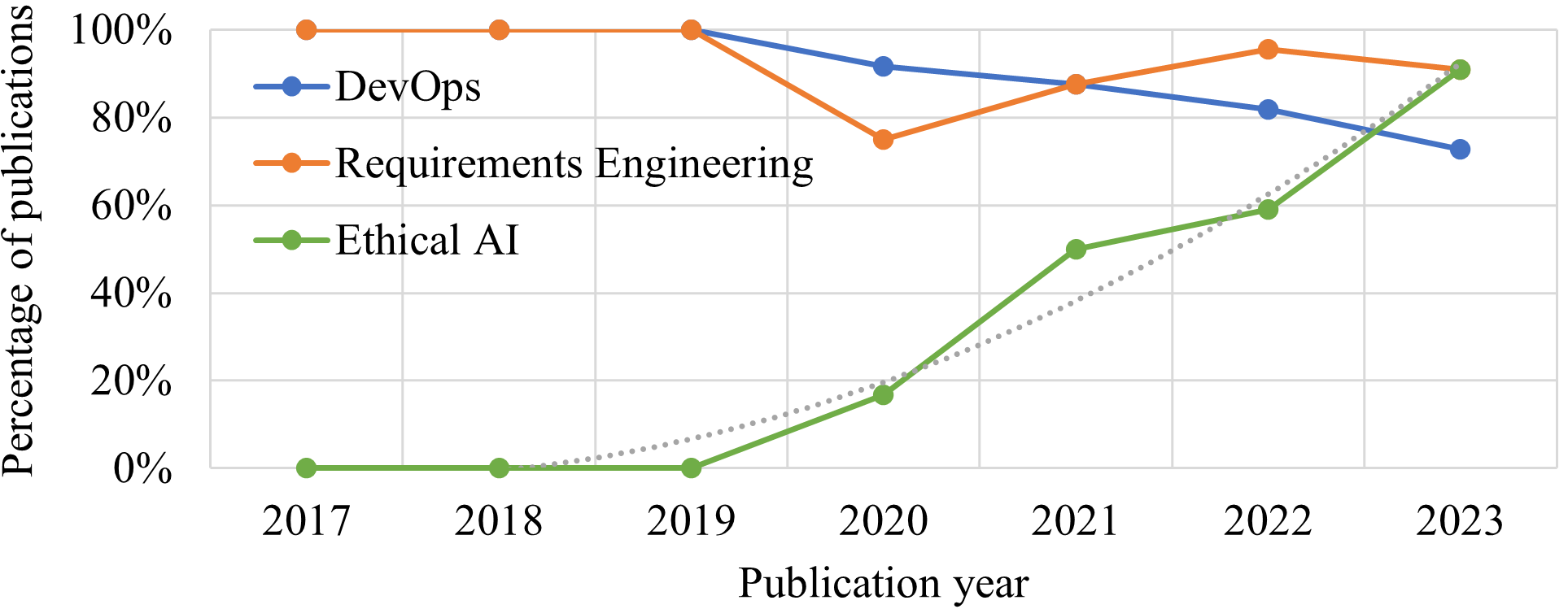}
    \caption{Prevalence of Ethical AI topics in the retrieved studies about AI and Agile.}
    \label{fig:evolutionethics}
\end{figure}

Finally, (vii) \textbf{ethical concerns about AI, as seen in Figure~\ref{fig:evolutionethics}, are significantly increasing their presence in the  literature since 2000}, especially in light of the rise of LLMs. As of now, there is little support for process automation and Agile development in ML-driven projects~\cite{Atouani2021}. As a result, traditional Agile practices may fail when used directly for ML application deployment~\cite{Shukla2022}.

\section{Artificial Intelligence to enhance Agile practices (RQ2)}

\begin{figure}
    \centering
    \includegraphics[width=.8\linewidth]{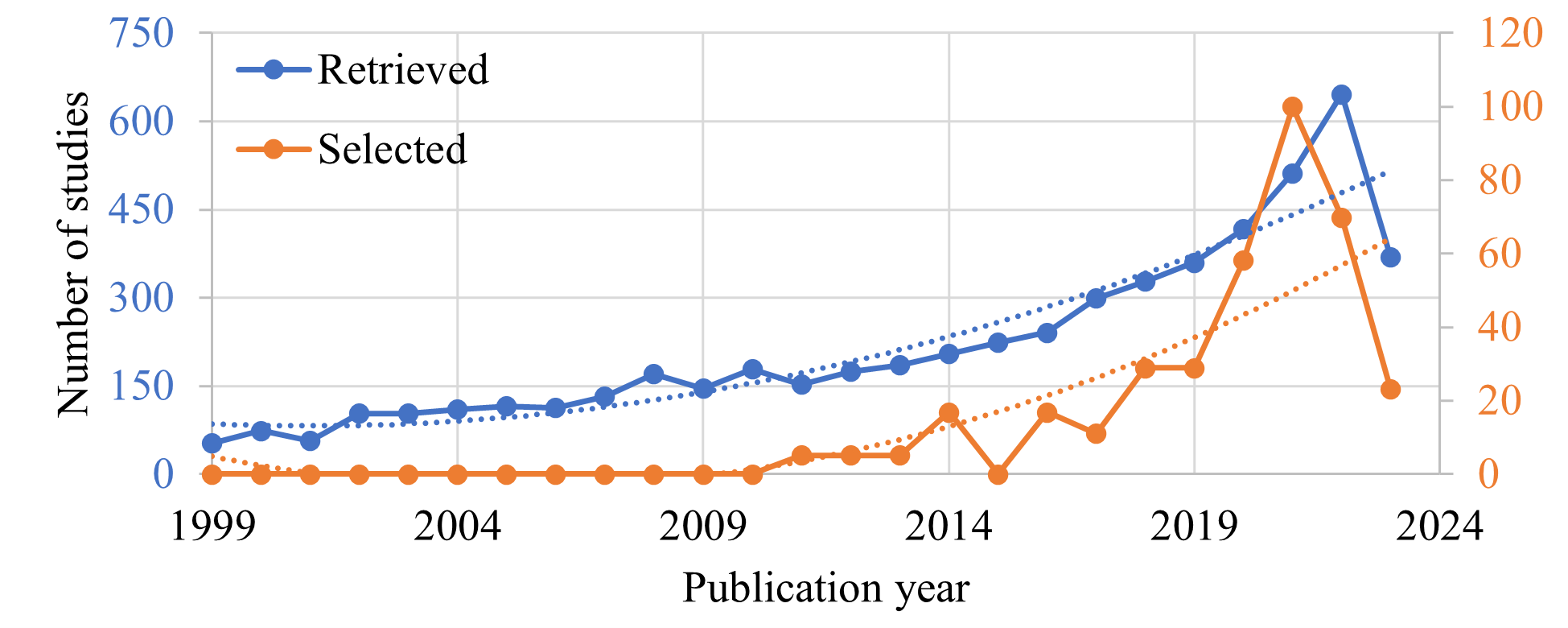}
    \caption{Evolution of the number of publications on AI-enhanced Agile processes retrieved in the initial search (blue) and after screening (orange) as of April 2023}
    \label{fig:AI4AgileTrend}
\end{figure}

We investigated the volume of studies that propose the use of Artificial Intelligence (AI) to assist in Agile software development practices, as described in Section~\ref{sec:method}. The evolution in time can be seen in Figure~\ref{fig:AI4AgileTrend}. Most of the increase can be explained by the apparent growing interest in applying AI and ML techniques to various domains, including software development. Nevertheless, it is interesting to note how the \textbf{more rapid increase of publications within the area of AI-assisted Agile practices (orange line in Figure~\ref{fig:AI4AgileTrend}) compared to papers covering AI and Agile-related topics (blue)}. 

AI can make many development processes more efficient. There are, however, tensions between the desire to deliver software quickly and the need for reliable products~\cite{rodriguez2017continuous}. DevOps approaches provide a pipeline to facilitate automation in continuous development and recent advances in the field of AI allow practitioners to integrate AI techniques within their DevOps process \textit{(M1)}, allowing scaling continuous and rapid development.

While test-driven development increases code quality and productivity, its industry adoption is low due to context-related factors~\cite{karac2018we}. Nevertheless, some of the limiting factors listed by Karac et al. (e.g., increased development time~\cite{karac2018we}) could be overcome by using automatic tools \textit{(M2)}: M{\i}s{\i}rl{\i} et al. stated in their 2011 paper that by inspecting less than a fourth of the code, it is possible to detect almost three fourths of software defects~\cite{misirli2011industrial}.
Nevertheless, regression testing needs for careful selection and prioritisation of test cases within test suites often too large to run entirely~\cite{bertolino2020learning}. Automatic regression test selection and prioritisation techniques \textit{(M3)} can reduce these testing efforts~\cite{Elsner2021,da2022machine} and increase the efficiency of continuous regression testing~\cite{marijan2020neural}. AI-enabled dynamic re-prioritisation of test cases throughout the development process further benefits developers~\cite{Katarzyna2021}.

AI can also help by identifying dependencies and conflicts in the requirements \textit{(M4)} and predicting the impact of changes to the backlog, allowing teams to make informed decisions about prioritisation \textit{(M5)}~\cite{rodriguez2017continuous,rahman2022challenges,gan2021automated,granlund2021mlops}. These tools can use ML algorithms to analyse various factors, such as the importance of a feature, the effort required to implement it, and the expected business value it will bring, to provide recommendations for prioritising user stories~\cite{kumar2022review,gan2021automated,pang2021promotion}.

\begin{table}[]
\centering
\caption{Challenges (RQ1) to mitigation strategies (RQ2) map}
\label{tab:mapping}
\resizebox{.85\textwidth}{!}{%
\begin{tabular}{p{1.5cm}p{1.5cm}p{9cm}}
\hline
\textbf{Chall.} & \textbf{Mitig.} & \textbf{Role of AI (RQ2)} \\
\hline
i & M4,5 & Identify dependencies and conflicts, prioritise \\
ii & M2,4 & Inform prioritisation decisions, indirectly reduce time needed \\
iii & M1,2,4 & Scaling continuous and rapid development in larger teams \\
iv & M1-5 & Automating tedious tasks, handling requirements dependencies, checking requirements' coverage and overlap \\
v & M4 & Clear responsibilities and stakeholders needs, communication \\
vi & M4-5 & Estimate impacts, prioritise commercial goals and value creation \\
vii & M3 & Re-evaluate AI systems whose performance might change \\
\hline
\end{tabular} }
\end{table}

Table~\ref{tab:mapping} maps these strategies to enhance Agile processes with AI \{M1,...,M5\} to the challenges \{i,...,vii\} identified in the retrieved literature. It also clarifies the role that AI plays in the mitigation of said challenges, addressing \textbf{RQ2}. 

\section{Challenges in AI-assisted Agile and trends (RQ3)}

Given there are only few tools for continuous deployment in ML-driven projects, Atouani et al. suggest a more unified framework that includes process automation and DevOps~\cite{Atouani2021}: MLOps, discusses the application of DevOps principles to ML systems but not the other way around. Some frameworks already \textbf{facilitate reflection within the development team on the development of Trustworthy AI software and helps developers integrate ethics} (increasingly receiving academic attention) into artefacts~\cite{AIact}. 

The adoption of AI in Agile software development processes brings new challenges related to security: e.g., cyber-attacks pose a critical threat, especially to safety-critical products that use deep neural networks~\cite{borg2018safely,menzies2012local} and LLMs~\cite{foundation}. Historically, obtaining appropriate data to train AI models has also been a challenge~\cite{menzies2012local,foundation} although recent research suggests that it is possible to combine software-related data sources, from different contexts or manufacturers, to improve predictions~\cite{menzies2012local,borg2018safely}. 
Despite these developments, it is important to remember that AI can receive adversarial attacks through data~\cite{borg2018safely}. 
Then, software security practices must be adopted to protect against attacks and ensure data privacy: requirements should be added, monitored, and prioritised~\cite{Krasteva2020,lariosvargas2022}. 

% Rodr{\'\i}guez et al. also mention the ``5-why'' practice to find the root cause of problems is one of the main benefits of lean thinking~\cite{rodriguez2019advances}. Recent trends in the literature show that Natural Language Processing and ontologies (often connected to user stories) are receiving increasing academic interest. Further research could therefore focus on AI for root cause analysis.

\section{Conclusion and Future Work}

There is a clear tendency to use AI to assist in daily tasks, including Agile processes, as discussed in Section~\ref{sec:results}. The integration of AI in Agile software development can help improve software development outcomes by enhancing accuracy, efficiency, and safety while reducing development time~\cite{lariosvargas2022,Boualouache2023}. However, there are also challenges associated with integrating AI and Agile methodologies, such as the need for specialised technical expertise, and integrating AI into Agile software development processes requires careful consideration of the context~\cite{karac2018we}. Additionally, the system's ability to adapt to changing contexts is also crucial, as Agile methodologies prioritise flexibility and adaptability to the ever-changing market~\cite{lariosvargas2022}. Without the ability to adapt to changing requirements, an AI system may not be suitable for integration with Agile methodologies at all. 

Some of the most critical challenges are related to human factors, e.g., ensuring that developers have the skills to work with AI and align expectations effectively~\cite{lariosvargas2022,AIact}. As we continue to explore the integration of AI with software development processes, it is important that we do not overlook human factors. To truly attain benefits, we must prioritise effective communication and understanding between human developers and AI-assistants~\cite{lariosvargas2022,AIact,yasolga2023}. By fostering strong collaboration between humans and machines, we can unlock new levels of productivity and innovation, also in software development. As such, further research on a number of human factors need to be conducted to fully understand the impact of AI on software development processes and software developers.

It is interesting to note that some of the most common tools for Agile software development practitioners and related search terms according to Google Trends\footnote{Visit \url{https://trends.google.com/}}, such as JIRA), rarely appear when scanning the academic literature. This might be interpreted as a gap in the existing literature that could be addressed in future work through empirical studies to investigate the \textit{actual} use of AI.

% \section*{Acknowledgements}
% Thanks to Prof. Berger and Assoc. Prof. Horkoff for their valuable guidance. This work was supported by the Vinnova project ASPECT [2021-04347].

%%%%%%%%%%%%%%%%%%%%%%%%%%%%%%%%%%%%%%%%%%%

\bibliographystyle{ieeetr}
\bibliography{sample-base}

\end{document}

\mainmatter              % start of a contribution
\title{AI-assisted Agile software development: a Systematic Literature Review - Appendix A}
\titlerunning{Appendix A}  % abbreviated title (for running head)
\author{submission 3799}
% \author{Ivar Ekeland\inst{1} \and Roger Temam\inst{2}
% Jeffrey Dean \and David Grove \and Craig Chambers \and Kim~B.~Bruce \and
% Elsa Bertino}
%
% \authorrunning{Ivar Ekeland et al.} % abbreviated author list (for running head)
% %
% %%%% list of authors for the TOC (use if author list has to be modified)
% \tocauthor{Ivar Ekeland, Roger Temam, Jeffrey Dean, David Grove,
% Craig Chambers, Kim B. Bruce, and Elisa Bertino}
% %
% \institute{Princeton University, Princeton NJ 08544, USA,\\
% \email{I.Ekeland@princeton.edu},\\ WWW home page:
% \texttt{http://users/\homedir iekeland/web/welcome.html}
% \and
% Universit\'{e} de Paris-Sud,
% Laboratoire d'Analyse Num\'{e}rique, B\^{a}timent 425,\\
% F-91405 Orsay Cedex, France}

\maketitle              % typeset the title of the contribution

%%%%%%%%%%%%%%%%%%%%%%%%%%%%%%%%%%%%%%%%%%%

\section{Study selection and data extraction}

The query used in order to search relevant papers in alternative databases was: \texttt{("AI-Assisted" OR "Artificial Intelligence" OR AI OR "role of AI" OR "machine learning" OR ML OR "AI-powered") AND ("agile software development" OR "backlog prioritization" OR "user Story Mapping" OR "Continuous Integration" OR "Sprint Planning" OR "Continuous Delivery")}.

From each of the selected papers, the information reported in Table~\ref{tab:dataextraction} was extracted. For the larger set of papers (as discussed in the main paper), only D1-5 were extracted.

\begin{table}[]
    \centering
    \caption{Data extracted from each study}
    \label{tab:dataextraction}
    % \resizebox{\linewidth}{!}{
    \begin{tabular}{lll}
    \hline
    \textbf{ID} & \textbf{Data item} & \textbf{RQ} \\
    \hline
    D1 & Author (only first) & Demographic data \\
    D2 & Title & Demographic data \\
    D3 & Venue & Demographic data \\
    D4 & Abstract & Demographic data \\
    D7 & Publication type & Demographic data \\
    D5 & Keywords & RQ1,2 \\
    D6 & Year & RQ2 \\
    D8 & Industry domain & RQ3 \\
    D7 & Frequent words & RQ3 \\
    \hline
    \end{tabular} %}
\end{table}

\section{Selected studies}

\noindent \textbf{ACM Digital Library}:
\begin{itemize}
    \item Research on the Application of Artificial Intelligence Algorithms in Drought Prediction (2023) \url{https://doi.org/10.1145/3573428.3573753}
    \item Security Patterns for Machine Learning: The Data-Oriented Stages (2023) \url{https://doi.org/10.1145/3551902.3565070}
    \item The AI Tech-Stack Model (2023) \url{https://doi.org/10.1145/3568026}
    \item Conceptual Challenges of Researching Artificial Intelligence in Public Administrations (2022) \url{https://doi.org/10.1145/3543434.3543441}
    \item Application of Artificial Intelligence in the Process of Rehabilitation of Mentally Ill Patients Returning to Society (2022) \url{https://doi.org/10.1145/3570773.3570803}
    \item Application of Artificial Intelligence in Mental Health and Mental Illnesses (2022) \url{https://doi.org/10.1145/3570773.3570834}
    \item Failure Prediction Using Transfer Learning in Large-Scale Continuous Integration Environments (2022) \url{https://dl.acm.org/doi/abs/10.5555/3566055.3566079}
    \item Current Status of the Application of Artificial Intelligence Technology in the Prevention and Control of Novel Infectious Coronavirus Pneumonia (2022) \url{https://doi.org/10.1145/3570773.3570810}
    \item HybridCISave: A Combined Build and Test Selection Approach in Continuous Integration (2022) \url{https://doi.org/10.1145/3576038}
    \item Predicting Build Outcomes in Continuous Integration Using Textual Analysis of Source Code Commits (2022) \url{https://doi.org/10.1145/3558489.3559070}
    \item Research on Development and Integration of Artificial Intelligence Aided Diagnosis System (2022) \url{https://doi.org/10.1145/3544109.3544178}
    \item Lessons from Eight Years of Operational Data from a Continuous Integration Service: An Exploratory Case Study of CircleCI (2022) \url{https://doi.org/10.1145/3510003.3510211}
    \item Workshop: Machine Learning in Software Quality (2022) \url{https://dl.acm.org/doi/abs/10.5555/3566055.3566095}
    \item Continuous Integration and Delivery Practices for Cyber-Physical Systems: An Interview-Based Study (2022) \url{https://doi.org/10.1145/3571854}
    \item Code Smells for Machine Learning Applications (2022) \url{https://doi.org/10.1145/3522664.3528620}
    \item On the Mechanism of Artificial Intelligence Affecting Sports Tourism Industry (2022) \url{https://doi.org/10.1145/3545822.3545831}
    \item Research on Automatic Composition Based on Multiple Machine Learning Models (2022) \url{https://doi.org/10.1145/3495018.3495366}
    \item Middleware 101: What to Know Now and for the Future (2022) \url{https://doi.org/10.1145/3526211}
    \item Middleware 101 (2022) \url{https://doi.org/10.1145/3546958}
    \item Industryâ€“Academia Research Collaboration and Knowledge Co-Creation: Patterns and Anti-Patterns (2022) \url{https://doi.org/10.1145/3494519}
    \item Predictability and Surprise in Large Generative Models (2022) \url{https://doi.org/10.1145/3531146.3533229}
    \item What is an AI engineer? An empirical analysis of job ads in The Netherlands (2022) \url{https://dl.acm.org/doi/abs/10.1145/3522664.3528594}
    \item Data sovereignty for AI pipelines: lessons learned from an industrial project at Mondragon corporation (2022) \url{https://dl.acm.org/doi/abs/10.1145/3522664.3528593}
    \item Research on Development and Integration of Artificial Intelligence Aided Diagnosis System (2022) \url{https://dl.acm.org/doi/abs/10.1145/3544109.3544178}
    \item Failure Prediction Using Machine Learning in IBM WebSphere Liberty Continuous Integration Environment (2021) \url{https://dl.acm.org/doi/abs/10.5555/3507788.3507798}
    \item Investigating Documented Information for Accurate Effort Estimation in Agile Software Development (2021) \url{https://doi.org/10.1145/3468264.3473106}
    \item A Pilot Study of Requirement Prioritization Techniques in Agile Software Development. (2021) \url{https://doi.org/10.1145/3494885.3494912}
    \item Investigating Continuous Delivery on IoT Systems (2021) \url{https://doi.org/10.1145/3493244.3493261}
    \item Continuous Delivery of Software on IoT Devices (2021) \url{https://doi.org/10.1109/MODELS-C.2019.00112}
    \item Reducing Cost in Continuous Integration with a Collection of Build Selection Approaches (2021) \url{https://doi.org/10.1145/3468264.3473103}
    \item Mlr3pipelines â€” Flexible Machine Learning Pipelines in R (2021) \url{https://www.jmlr.org/papers/volume22/21-0281/21-0281.pdf}
    \item Dynamic Time Window Based Reward for Reinforcement Learning in Continuous Integration Testing (2021) \url{https://doi.org/10.1145/3457913.3457930}
    \item Empirically Evaluating Readily Available Information for Regression Test Optimization in Continuous Integration (2021) \url{https://doi.org/10.1145/3460319.3464834}
    \item Artifact and Reference Models for Generative Machine Learning Frameworks and Build Systems (2021) \url{https://doi.org/10.1145/3486609.3487199}
    \item Gegelati: Lightweight Artificial Intelligence through Generic and Evolvable Tangled Program Graphs (2021) \url{https://doi.org/10.1145/3441110.3441575}
    \item The SPACE of Developer Productivity: There's More to It than You Think. (2021) \url{https://doi.org/10.1145/3454122.3454124}
    \item Static Analysis: An Introduction: The Fundamental Challenge of Software Engineering is One of Complexity. (2021) \url{https://doi.org/10.1145/3487019.3487021}
    \item Static Analysis (2021) \url{https://doi.org/10.1145/3486592}
    \item Software Development in Disruptive Times (2021) \url{https://doi.org/10.1145/3453932}
    \item A Survey of Flaky Tests (2021) \url{https://doi.org/10.1145/3476105}
    \item Serverless Edge Computing: Vision and Challenges (2021) \url{https://doi.org/10.1145/3437378.3444367}
    \item Understanding Software-2.0: A Study of Machine Learning Library Usage and Evolution (2021) \url{https://doi.org/10.1145/3453478}
    \item When and How to Make Breaking Changes: Policies and Practices in 18 Open Source Software Ecosystems (2021) \url{https://doi.org/10.1145/3447245}
    \item A Survey on Automated Log Analysis for Reliability Engineering (2021) \url{https://doi.org/10.1145/3460345}
    \item Serverless Edge Computing: Vision and Challenges (2021) \url{https://doi.org/10.1145/3437378.3444367}
    \item Building Continuous Integration Services for Machine Learning (2020) \url{https://doi.org/10.1145/3394486.3403290}
    \item Tackling Build Failures in Continuous Integration (2020) \url{https://doi.org/10.1109/ASE.2019.00150}
    \item The State of the ML-Universe: 10 Years of Artificial Intelligence \&amp; Machine Learning Software Development on GitHub (2020) \url{https://doi.org/10.1145/3379597.3387473}
    \item Enabling Continuous Improvement of a Continuous Integration Process (2020) \url{https://doi.org/10.1109/ASE.2019.00151}
    \item On the Influence of Different Perspectives on Evaluating the Teamwork Quality in the Context of Agile Software Development (2020) \url{https://doi.org/10.1145/3422392.3422397}
    \item An Automatic Artificial Intelligence Training Platform Based on Kubernetes (2020) \url{https://doi.org/10.1145/3378904.3378921}
    \item A Cost-Efficient Approach to Building in Continuous Integration (2020) \url{https://doi.org/10.1145/3377811.3380437}
    \item Implications of Resurgence in Artificial Intelligence for Research Collaborations in Software Engineering (2020) \url{https://doi.org/10.1145/3356773.3356813}
    \item Tslearn, a Machine Learning Toolkit for Time Series Data (2020) \url{https://jmlr.org/papers/volume21/20-091/20-091.pdf}
    \item The Use of Change Point Detection to Identify Software Performance Regressions in a Continuous Integration System (2020) \url{https://doi.org/10.1145/3358960.3375791}
    \item APL since 1978 (2020) \url{https://doi.org/10.1145/3386319}
    \item Re-Examining Whether, Why, and How Human-AI Interaction Is Uniquely Difficult to Design (2020) \url{https://doi.org/10.1145/3313831.3376301}
    \item A History of MATLAB (2020) \url{https://doi.org/10.1145/3386331}
    \item How Enterprises Adopt Agile Forms of Organizational Design: A Multiple-Case Study (2020) \url{https://doi.org/10.1145/3380799.3380807}
    \item MP-SPDZ: A Versatile Framework for Multi-Party Computation (2020) \url{https://doi.org/10.1145/3372297.3417872}
    \item Re-Examining Whether, Why, and How Human-AI Interaction Is Uniquely Difficult to Design (2020) \url{https://doi.org/10.1145/3313831.3376301}
    \item MP-SPDZ: A Versatile Framework for Multi-Party Computation (2020) \url{https://doi.org/10.1145/3372297.3417872}
    \item CC2Vec: Distributed Representations of Code Changes (2020) \url{https://doi.org/10.1145/3377811.3380361}
    \item What is the Vocabulary of Flaky Tests? (2020) \url{https://doi.org/10.1145/3379597.3387482}
    \item Learning-to-rank vs ranking-to-learn: Strategies for regression testing in continuous integration (2020) \url{https://dl.acm.org/doi/abs/10.1145/3377811.3380369}
    \item Towards Effective AI-Powered Agile Project Management (2019) \url{https://doi.org/10.1109/ICSE-NIER.2019.00019}
    \item A Novel Framework for Change Requirement Management (CRM) In Agile Software Development (ASD) (2019) \url{https://doi.org/10.1145/3357419.3357438}
    \item Software Engineering for Machine Learning: A Case Study (2019) \url{https://doi.org/10.1109/ICSE-SEIP.2019.00042}
    \item A New Golden Age for Computer Architecture (2019) \url{https://doi.org/10.1145/3282307}
    \item Keeping Master Green at Scale (2019) \url{https://doi.org/10.1145/3302424.3303970}
    \item The Reliability of Enterprise Applications: Understanding Enterprise Reliability (2019) \url{https://doi.org/10.1145/3371595.3374665}
    \item DAML: The Contract Language of Distributed Ledgers (2019) \url{https://doi.org/10.1145/3343046}
    \item OpenPiton: An Open Source Hardware Platform for Your Research (2019) \url{https://doi.org/10.1145/3366343}
    \item The Reliability of Enterprise Applications (2019) \url{https://doi.org/10.1145/3369756}
    \item Repairnator Patches Programs Automatically (2019) \url{https://doi.org/10.1145/3349589}
    \item Robust Log-Based Anomaly Detection on Unstable Log Data (2019) \url{https://doi.org/10.1145/3338906.3338931}
    \item Semantic Fuzzing with Zest (2019) \url{https://doi.org/10.1145/3293882.3330576}
    \item A New Golden Age for Computer Architecture (2019) \url{https://doi.org/10.1145/3282307}
    \item Robust Log-Based Anomaly Detection on Unstable Log Data (2019) \url{https://doi.org/10.1145/3338906.3338931}
    \item Software Engineering for Machine Learning: A Case Study (2019) \url{https://doi.org/10.1109/ICSE-SEIP.2019.00042}
    \item Understanding Flaky Tests: The Developerâ€™s Perspective (2019) \url{https://doi.org/10.1145/3338906.3338945}
    \item Semantic Fuzzing with Zest (2019) \url{https://doi.org/10.1145/3293882.3330576}
    \item Toward an Open-Source Digital Flow: First Learnings from the OpenROAD Project (2019) \url{https://doi.org/10.1145/3316781.3326334}
    \item IFixR: Bug Report Driven Program Repair (2019) \url{https://doi.org/10.1145/3338906.3338935}
    \item Sentiment analysis for software engineering: How far can we go? (2018) \url{https://dl.acm.org/doi/abs/10.1145/3180155.3180195}
    \item ACONA: active online model adaptation for predicting continuous integration build failures (2018) \url{https://dl.acm.org/doi/abs/10.1145/3183440.3195012}
    \item Software Analytics in Continuous Delivery: A Case Study on Success Factors (2018) \url{https://doi.org/10.1145/3239235.3240505}
    \item Lessons from Building Static Analysis Tools at Google (2018) \url{https://doi.org/10.1145/3188720}
    \item Service Fabric: A Distributed Platform for Building Microservices in the Cloud (2018) \url{https://doi.org/10.1145/3190508.3190546}
    \item Contextual Understanding of Microservice Architecture: Current and Future Directions (2018) \url{https://doi.org/10.1145/3183628.3183631}
    \item Architectural Principles for Cloud Software (2018) \url{https://doi.org/10.1145/3104028}
    \item Lessons from Building Static Analysis Tools at Google (2018) \url{https://doi.org/10.1145/3188720}
    \item Contextual Understanding of Microservice Architecture: Current and Future Directions (2018) \url{https://doi.org/10.1145/3183628.3183631}
    \item Sentiment Analysis for Software Engineering: How Far Can We Go? (2018) \url{https://doi.org/10.1145/3180155.3180195}
    \item Neural-Machine-Translation-Based Commit Message Generation: How Far Are We? (2018) \url{https://doi.org/10.1145/3238147.3238190}
    \item The Power of Bots: Characterizing and Understanding Bots in OSS Projects (2018) \url{https://doi.org/10.1145/3274451}
    \item Architectural Principles for Cloud Software (2018) \url{https://doi.org/10.1145/3104028}
    \item Continuous Reasoning: Scaling the Impact of Formal Methods (2018) \url{https://doi.org/10.1145/3209108.3209109}
    \item An Agile Software Engineering Method to Design Blockchain Applications (2018) \url{https://doi.org/10.1145/3290621.3290627}
    \item FAST Approaches to Scalable Similarity-Based Test Case Prioritization (2018) \url{https://doi.org/10.1145/3180155.3180210}
    \item HireBuild: An Automatic Approach to History-Driven Repair of Build Scripts (2018) \url{https://doi.org/10.1145/3180155.3180181}
    \item Redefining Prioritization: Continuous Prioritization for Continuous Integration (2018) \url{https://doi.org/10.1145/3180155.3180213}
    \item Reinforcement Learning for Automatic Test Case Prioritization and Selection in Continuous Integration (2017) \url{https://doi.org/10.1145/3092703.3092709}
    \item Impact of Continuous Integration on Code Reviews (2017) \url{https://doi.org/10.1109/MSR.2017.39}
    \item Reinforcement Learning for Automatic Test Case Prioritization and Selection in Continuous Integration (2017) \url{https://doi.org/10.1145/3092703.3092709}
    \item Taming Google-Scale Continuous Testing (2017) \url{https://doi.org/10.1109/ICSE-SEIP.2017.16}
    \item Automated Identification of Security Issues from Commit Messages and Bug Reports (2017) \url{https://doi.org/10.1145/3106237.3117771}
    \item A Meta-Analysis of Pair-Programming in Computer Programming Courses: Implications for Educational Practice (2017) \url{https://doi.org/10.1145/2996201}
    \item Sieve: Actionable Insights from Monitored Metrics in Distributed Systems (2017) \url{https://doi.org/10.1145/3135974.3135977}
    \item Automatically Measuring the Maintainability of Service- and Microservice-Based Systems: A Literature Review (2017) \url{https://doi.org/10.1145/3143434.3143443}
    \item API code recommendation using statistical learning from fine-grained changes (2016) \url{https://dl.acm.org/doi/abs/10.1145/2950290.2950333}
    \item Usage, Costs, and Benefits of Continuous Integration in Open-Source Projects (2016) \url{https://doi.org/10.1145/2970276.2970358}
    \item Measuring and Understanding Team Development by Capturing Self-Assessed Enthusiasm and Skill Levels (2016) \url{https://doi.org/10.1145/2791394}
    \item Usage, Costs, and Benefits of Continuous Integration in Open-Source Projects (2016) \url{https://doi.org/10.1145/2970276.2970358}
    \item Using (Bio)Metrics to Predict Code Quality Online (2016) \url{https://doi.org/10.1145/2884781.2884803}
    \item Usage, Costs, and Benefits of Continuous Integration in Open-Source Projects (2016) \url{https://doi.org/10.1145/2970276.2970358}
    \item Work Practices and Challenges in Pull-Based Development: The Contributor's Perspective (2016) \url{https://doi.org/10.1145/2884781.2884826}
    \item API Code Recommendation Using Statistical Learning from Fine-Grained Changes (2016) \url{https://doi.org/10.1145/2950290.2950333}
    \item An Empirical Investigation into the Nature of Test Smells (2016) \url{https://doi.org/10.1145/2970276.2970340}
    \item Madmom: A New Python Audio and Music Signal Processing Library (2016) \url{https://doi.org/10.1145/2964284.2973795}
    \item Using (Bio)Metrics to Predict Code Quality Online (2016) \url{https://doi.org/10.1145/2884781.2884803}
    \item Continuous Delivery of Composite Solutions: A Case for Collaborative Software Defined PaaS Environments (2015) \url{https://doi.org/10.1145/2756594.2756595}
    \item IDyLL: Indoor Localization Using Inertial and Light Sensors on Smartphones (2015) \url{https://doi.org/10.1145/2750858.2807540}
    \item Effort Estimation in Agile Software Development: A Systematic Literature Review (2014) \url{https://doi.org/10.1145/2639490.2639503}
    \item Undergraduate Software Engineering: Addressing the Needs of Professional Software Development: Addressing the Needs of Professional Software Development (2014) \url{https://doi.org/10.1145/2636163.2653382}
    \item Undergraduate Software Engineering (2014) \url{https://doi.org/10.1145/2632361}
    \item Effort Estimation in Agile Software Development: A Systematic Literature Review (2014) \url{https://doi.org/10.1145/2639490.2639503}
    \item Fine-Grained and Accurate Source Code Differencing (2014) \url{https://doi.org/10.1145/2642937.2642982}
    \item Learning Natural Coding Conventions (2014) \url{https://doi.org/10.1145/2635868.2635883}
    \item Cowboys, Ankle Sprains, and Keepers of Quality: How is Video Game Development Different from Software Development? (2014) \url{https://doi.org/10.1145/2568225.2568226}
    \item Use It Free: Instantly Knowing Your Phone Attitude (2014) \url{https://doi.org/10.1145/2639108.2639110}
    \item Parameter-Less Population Pyramid (2014) \url{https://doi.org/10.1145/2576768.2598350}
    \item Fine-Grained and Accurate Source Code Differencing (2014) \url{https://doi.org/10.1145/2642937.2642982}
    \item Learning Natural Coding Conventions (2014) \url{https://doi.org/10.1145/2635868.2635883}
    \item Use It Free: Instantly Knowing Your Phone Attitude (2014) \url{https://doi.org/10.1145/2639108.2639110}
    \item Effort Estimation in Agile Software Development: A Systematic Literature Review (2014) \url{https://doi.org/10.1145/2639490.2639503}
    \item Cowboys, Ankle Sprains, and Keepers of Quality: How is Video Game Development Different from Software Development? (2014) \url{https://doi.org/10.1145/2568225.2568226}
    \item Parameter-Less Population Pyramid (2014) \url{https://doi.org/10.1145/2576768.2598350}
    \item A Pattern Language for Inter-Team Knowledge Sharing in Agile Software Development (2013) \url{https://hillside.net/plop/2013/papers/Group4/plop13_preprint_36.pdf}
    \item Automated Root Cause Isolation of Performance Regressions during Software Development (2013) \url{https://doi.org/10.1145/2479871.2479879}
    \item Machine Learning and Algorithms; Agile Development (2012) \url{https://doi.org/10.1145/2240236.2240239}
    \item Systematic Literature Studies: Database Searches vs. Backward Snowballing (2012) \url{https://doi.org/10.1145/2372251.2372257}
    \item Machine Learning and Algorithms; Agile Development (2012) \url{https://doi.org/10.1145/2240236.2240239}
    \item Test Driven Development: The State of the Practice (2012) \url{https://doi.org/10.1145/2184512.2184550}
    \item Systematic Literature Studies: Database Searches vs. Backward Snowballing (2012) \url{https://doi.org/10.1145/2372251.2372257}
    \item Method-Level Bug Prediction (2012) \url{https://doi.org/10.1145/2372251.2372285}
    \item Rafting the Agile Waterfall: Value Based Conflicts of Agile Software Development (2011) \url{https://doi.org/10.1145/2396716.2396731}
    \item B.Y.O.C (1,342 Times and Counting) (2011) \url{https://doi.org/10.1145/1897852.1897870}
    \item Evolving Patches for Software Repair (2011) \url{https://doi.org/10.1145/2001576.2001768}
    \item An Empirical Study on the Relationship between the Use of Agile Practices and the Success of Scrum Projects (2010) \url{https://doi.org/10.1145/1852786.1852835}
    \item Biharmonic Distance (2010) \url{https://doi.org/10.1145/1805964.1805971}
    \item Studying Communication in Agile Software Development: A Research Framework and Pilot Study (2009) \url{https://doi.org/10.1145/1641389.1641394}
    \item Process-Centered Review of Object Oriented Software Development Methodologies (2008) \url{https://doi.org/10.1145/1322432.1322435}
    \item Process-Centered Review of Object Oriented Software Development Methodologies (2008) \url{https://doi.org/10.1145/1322432.1322435}
    \item Building Collaboration into IDEs: Edit\&gt;Compile\&gt;Run\&gt;Debug\&gt;Collaborate? (2003) \url{https://doi.org/10.1145/966789.966803}
    \item Building Collaboration into IDEs: Edit\&gt;Compile\&gt;Run\&gt;Debug\&gt;Collaborate? (2003) \url{https://doi.org/10.1145/966789.966803}
\end{itemize}

\noindent \textbf{ArXiV}:
\begin{itemize}
    \item A Survey on Machine Learning-based Misbehavior Detection Systems for 5G and Beyond Vehicular Networks (2023) \url{https://arxiv.org/pdf/2201.10500.pdf}
    \item Towards Human-Bot Collaborative Software Architecting with ChatGPT (2023) \url{https://arxiv.org/pdf/2302.14600.pdf}
    \item DASP: A Framework for Driving the Adoption of Software Security Practices (2022) \url{https://arxiv.org/pdf/2205.12388.pdf}
    \item On the opportunities and risks of foundation models (2021) \url{https://arxiv.org/pdf/2108.07258.pdf}
    \item Safely entering the deep: A review of verification and validation for machine learning and a challenge elicitation in the automotive industry (2018) \url{https://arxiv.org/pdf/1812.05389.pdf}
\end{itemize}

\noindent \textbf{Google Scholar}:
\begin{itemize}
    \item Artificial Intelligence Enables Agile Software Development Life Cycle (2023) \url{https://scholar.googleusercontent.com/scholar.bib?q=info:n76UqsnpBOwJ:scholar.google.com/\&amp;output=citation\&amp;scisdr=CgXSgnCxEMTkn4zZZMc:AAGBfm0AAAAAZBjffMZ5GAp9IF_DRofs2ecgV6FcEgb0\&amp;scisig=AAGBfm0AAAAAZBjffHC6cYPypoRDsBBEtScBMjvWIG5j\&amp;scisf=4\&amp;ct=citation\&amp;cd=2\&amp;hl=en}
    \item Comparative Study of Machine Learning Test Case Prioritization for Continuous Integration Testing (2022) \url{https://scholar.googleusercontent.com/scholar.bib?q=info:n4HENcf522kJ:scholar.google.com/\&amp;output=citation\&amp;scisdr=CgXSgnCxEMTkn4zZZMc:AAGBfm0AAAAAZBjffMZ5GAp9IF_DRofs2ecgV6FcEgb0\&amp;scisig=AAGBfm0AAAAAZBjffDuEcbSPvIPOKe_qxXhGHmBeSH7A\&amp;scisf=4\&amp;ct=citation\&amp;cd=4\&amp;hl=en}
    \item Machine Learning Regression Techniques for Test Case Prioritization in Continuous Integration Environment (2022) \url{https://scholar.googleusercontent.com/scholar.bib?q=info:rSuA4TwTB2EJ:scholar.google.com/\&amp;output=citation\&amp;scisdr=CgXSgnCxEMTkn4zZZMc:AAGBfm0AAAAAZBjffMZ5GAp9IF_DRofs2ecgV6FcEgb0\&amp;scisig=AAGBfm0AAAAAZBjffHdtsRLmBEcEHrLcwyiGdHzkM4QM\&amp;scisf=4\&amp;ct=citation\&amp;cd=5\&amp;hl=en}
    \item Reinforcement Learning Reward Function for Test Case Prioritization in Continuous Integration (2022) \url{https://scholar.googleusercontent.com/scholar.bib?q=info:bIF9qIFV-oYJ:scholar.google.com/\&amp;output=citation\&amp;scisdr=CgXSgnCxEMTkn4zZZMc:AAGBfm0AAAAAZBjffMZ5GAp9IF_DRofs2ecgV6FcEgb0\&amp;scisig=AAGBfm0AAAAAZBjffGZyYqHcWTYb1Xj0J46f2bNMtMOb\&amp;scisf=4\&amp;ct=citation\&amp;cd=11\&amp;hl=en}
    \item Arcalog: Enhancing Continuous Integration Systems with Assisted Root Cause Analysis (2022) \url{https://scholar.googleusercontent.com/scholar.bib?q=info:KLd2khk3desJ:scholar.google.com/\&amp;output=citation\&amp;scisdr=CgXSgnCxEMTkn4zZZMc:AAGBfm0AAAAAZBjffMZ5GAp9IF_DRofs2ecgV6FcEgb0\&amp;scisig=AAGBfm0AAAAAZBjffGE39FP7fTu1AhCheKzTKE5lcxF-\&amp;scisf=4\&amp;ct=citation\&amp;cd=18\&amp;hl=en}
    \item A Fuzzy AHP-based approach for prioritization of cost overhead factors in agile software development (2022) \url{https://scholar.googleusercontent.com/scholar.bib?q=info:Ye_xeEY5BVUJ:scholar.google.com/\&amp;output=citation\&amp;scisdr=CgXSgnCxEMTkn4zmv3w:AAGBfm0AAAAAZBjgp30tUya8c9UNvXjmOFbxF9EIs_Jp\&amp;scisig=AAGBfm0AAAAAZBjgpxwdgdZOOAGhMoXnNEaZ5ZnD8XX0\&amp;scisf=4\&amp;ct=citation\&amp;cd=27\&amp;hl=en}
    \item A Fuzzy AHP-based approach for prioritization of cost overhead factors in agile software development (2022) \url{https://scholar.googleusercontent.com/scholar.bib?q=info:JuJ0BywQSN0J:scholar.google.com/\&amp;output=citation\&amp;scisdr=CgXSgnCxEMTkn4zmv3w:AAGBfm0AAAAAZBjgp30tUya8c9UNvXjmOFbxF9EIs_Jp\&amp;scisig=AAGBfm0AAAAAZBjgp2zb07slXxxxRgPgepe2NphnOJeR\&amp;scisf=4\&amp;ct=citation\&amp;cd=28\&amp;hl=en}
    \item Challenging the Artifacts and Practices Adopted in Agile Software Development (2022) \url{https://scholar.googleusercontent.com/scholar.bib?q=info:87H1e7_ZSDkJ:scholar.google.com/\&amp;output=citation\&amp;scisdr=CgXSgnCxEMTkn4zmv3w:AAGBfm0AAAAAZBjgp30tUya8c9UNvXjmOFbxF9EIs_Jp\&amp;scisig=AAGBfm0AAAAAZBjgp0GSpsZmrQ3M0YoAaH8sl-a9WTRW\&amp;scisf=4\&amp;ct=citation\&amp;cd=29\&amp;hl=en}
    \item Machine learning to support code reviews in continuous integration (2021) \url{https://scholar.googleusercontent.com/scholar.bib?q=info:OxNSvI9y8AIJ:scholar.google.com/\&amp;output=citation\&amp;scisdr=CgXSgnCxEMTkn4zZZMc:AAGBfm0AAAAAZBjffMZ5GAp9IF_DRofs2ecgV6FcEgb0\&amp;scisig=AAGBfm0AAAAAZBjffGSn0MYDtmZOivm6j_GzUU81eKKg\&amp;scisf=4\&amp;ct=citation\&amp;cd=0\&amp;hl=en}
    \item Supervised Learning for Test Suit Selection in Continuous Integration (2021) \url{https://scholar.googleusercontent.com/scholar.bib?q=info:KiUzJ_CnEJYJ:scholar.google.com/\&amp;output=citation\&amp;scisdr=CgXSgnCxEMTkn4zZZMc:AAGBfm0AAAAAZBjffMZ5GAp9IF_DRofs2ecgV6FcEgb0\&amp;scisig=AAGBfm0AAAAAZBjffLSnlqqHlomBN_FfxFP7FbAfm50U\&amp;scisf=4\&amp;ct=citation\&amp;cd=7\&amp;hl=en}
    \item Failure prediction using machine learning in IBM WebSphere liberty continuous integration environment (2021) \url{https://scholar.googleusercontent.com/scholar.bib?q=info:OgnV8LstcUcJ:scholar.google.com/\&amp;output=citation\&amp;scisdr=CgXSgnCxEMTkn4zZZMc:AAGBfm0AAAAAZBjffMZ5GAp9IF_DRofs2ecgV6FcEgb0\&amp;scisig=AAGBfm0AAAAAZBjffHm_2YUx1k9nzbmKXfuDDBoe47vX\&amp;scisf=4\&amp;ct=citation\&amp;cd=8\&amp;hl=en}
    \item Engineering MLOps: Rapidly build, test, and manage production-ready machine learning life cycles at scale (2021) \url{https://scholar.googleusercontent.com/scholar.bib?q=info:WhvIbbaoqvQJ:scholar.google.com/\&amp;output=citation\&amp;scisdr=CgXSgnCxEMTkn4zZZMc:AAGBfm0AAAAAZBjffMZ5GAp9IF_DRofs2ecgV6FcEgb0\&amp;scisig=AAGBfm0AAAAAZBjffAxu-j1bBL76w3DFPUA5luj9twli\&amp;scisf=4\&amp;ct=citation\&amp;cd=14\&amp;hl=en}
    \item Success factors when transitioning to continuous deployment in software-intensive embedded systems (2021) \url{https://scholar.googleusercontent.com/scholar.bib?q=info:FrB4gJY-vBkJ:scholar.google.com/\&amp;output=citation\&amp;scisdr=CgXSgnCxEMTkn4zmv3w:AAGBfm0AAAAAZBjgp30tUya8c9UNvXjmOFbxF9EIs_Jp\&amp;scisig=AAGBfm0AAAAAZBjgp53g0PY5sFYSGR7AEcOqWB5Oo82y\&amp;scisf=4\&amp;ct=citation\&amp;cd=24\&amp;hl=en}
    \item Experimental Research on a Continuous Integrating pipeline with a Machine Learning approach: Master Thesis done in collaboration with Electronic Arts (2021) \url{https://scholar.googleusercontent.com/scholar.bib?q=info:JAEupJzdyR0J:scholar.google.com/\&amp;output=citation\&amp;scisdr=CgXSgnCxEMTkn4zmv3w:AAGBfm0AAAAAZBjgp30tUya8c9UNvXjmOFbxF9EIs_Jp\&amp;scisig=AAGBfm0AAAAAZBjgp-rwbheWIfH-KCYhqBvkO53GbkiU\&amp;scisf=4\&amp;ct=citation\&amp;cd=26\&amp;hl=en}
    \item Towards regulatory-compliant MLOps: Oravizioâ€™s journey from a machine learning experiment to a deployed certified medical product (2021) \url{https://scholar.googleusercontent.com/scholar.bib?q=info:9fo9ATOH1AkJ:scholar.google.com/\&amp;output=citation\&amp;scisdr=CgXSgnCxEMTkn4zmv3w:AAGBfm0AAAAAZBjgp30tUya8c9UNvXjmOFbxF9EIs_Jp\&amp;scisig=AAGBfm0AAAAAZBjgp6sRr_MG2UjI1yxZ-EwZIIkgunka\&amp;scisf=4\&amp;ct=citation\&amp;cd=30\&amp;hl=en}
    \item Machine learning models to predict agile methodology adoption (2020) \url{https://scholar.googleusercontent.com/scholar.bib?q=info:CGu7hOC5lcgJ:scholar.google.com/\&amp;output=citation\&amp;scisdr=CgXSgnCxEMTkn4zZZMc:AAGBfm0AAAAAZBjffMZ5GAp9IF_DRofs2ecgV6FcEgb0\&amp;scisig=AAGBfm0AAAAAZBjffGeYkkYiwG_q0RIEEZSDNCe-ybe_\&amp;scisf=4\&amp;ct=citation\&amp;cd=3\&amp;hl=en}
    \item Applying DevOps practices of continuous automation for machine learning (2020) \url{https://scholar.googleusercontent.com/scholar.bib?q=info:WGRrJkHKkokJ:scholar.google.com/\&amp;output=citation\&amp;scisdr=CgXSgnCxEMTkn4zZZMc:AAGBfm0AAAAAZBjffMZ5GAp9IF_DRofs2ecgV6FcEgb0\&amp;scisig=AAGBfm0AAAAAZBjffKRZVALdfwwDvZpumXjdesRN-wUA\&amp;scisf=4\&amp;ct=citation\&amp;cd=9\&amp;hl=en}
    \item An update on effort estimation in agile software development: A systematic literature review (2020) \url{https://scholar.googleusercontent.com/scholar.bib?q=info:emsQut0MofQJ:scholar.google.com/\&amp;output=citation\&amp;scisdr=CgXSgnCxEMTkn4zZZMc:AAGBfm0AAAAAZBjffMZ5GAp9IF_DRofs2ecgV6FcEgb0\&amp;scisig=AAGBfm0AAAAAZBjffNztqSJ8k8SYPqmC3UUEsZwumlJ8\&amp;scisf=4\&amp;ct=citation\&amp;cd=12\&amp;hl=en}
    \item An Automatic Artificial Intelligence Training Platform Based on Kubernetes (2020) \url{https://scholar.googleusercontent.com/scholar.bib?q=info:hu092NH6eWgJ:scholar.google.com/\&amp;output=citation\&amp;scisdr=CgXSgnCxEMTkn4zZZMc:AAGBfm0AAAAAZBjffMZ5GAp9IF_DRofs2ecgV6FcEgb0\&amp;scisig=AAGBfm0AAAAAZBjffL3mNgP1kMVI20EV7JaFaXN5lpMQ\&amp;scisf=4\&amp;ct=citation\&amp;cd=16\&amp;hl=en}
    \item A technique for transitioning of plan driven software development method to distributed agile software development (2020) \url{https://scholar.googleusercontent.com/scholar.bib?q=info:TlImF5Sk20gJ:scholar.google.com/\&amp;output=citation\&amp;scisdr=CgXSgnCxEMTkn4zZZMc:AAGBfm0AAAAAZBjffMZ5GAp9IF_DRofs2ecgV6FcEgb0\&amp;scisig=AAGBfm0AAAAAZBjffBT1eQdUuMG-qzz2jK73i0GQti9F\&amp;scisf=4\&amp;ct=citation\&amp;cd=17\&amp;hl=en}
    \item Adopting Agile Software Development methodologies in big data projects--a systematic literature review of experience reports (2020) \url{https://scholar.googleusercontent.com/scholar.bib?q=info:petirE9_PPEJ:scholar.google.com/\&amp;output=citation\&amp;scisdr=CgXSgnCxEMTkn4zmv3w:AAGBfm0AAAAAZBjgp30tUya8c9UNvXjmOFbxF9EIs_Jp\&amp;scisig=AAGBfm0AAAAAZBjgp8Eu8wNEUK0RWCzvrPoNyc9GE0nU\&amp;scisf=4\&amp;ct=citation\&amp;cd=20\&amp;hl=en}
    \item Test case selection and prioritization in continuous integration environment (2019) \url{https://scholar.googleusercontent.com/scholar.bib?q=info:n4yGRYB8mk8J:scholar.google.com/\&amp;output=citation\&amp;scisdr=CgXSgnCxEMTkn4zZZMc:AAGBfm0AAAAAZBjffMZ5GAp9IF_DRofs2ecgV6FcEgb0\&amp;scisig=AAGBfm0AAAAAZBjffGU3TMgvwQRdAAUdl84-c-UzSd-S\&amp;scisf=4\&amp;ct=citation\&amp;cd=13\&amp;hl=en}
    \item Predictive test selection (2019) \url{https://scholar.googleusercontent.com/scholar.bib?q=info:FaWx334BAP4J:scholar.google.com/\&amp;output=citation\&amp;scisdr=CgXSgnCxEMTkn4zZZMc:AAGBfm0AAAAAZBjffMZ5GAp9IF_DRofs2ecgV6FcEgb0\&amp;scisig=AAGBfm0AAAAAZBjffGT2_hRClyohxQajQ5GLLj_dvVzQ\&amp;scisf=4\&amp;ct=citation\&amp;cd=15\&amp;hl=en}
    \item Continuous integration logic (2019) \url{https://scholar.googleusercontent.com/scholar.bib?q=info:lQuPhNFAsKwJ:scholar.google.com/\&amp;output=citation\&amp;scisdr=CgXSgnCxEMTkn4zZZMc:AAGBfm0AAAAAZBjffMZ5GAp9IF_DRofs2ecgV6FcEgb0\&amp;scisig=AAGBfm0AAAAAZBjffBaD1oZsK-xkgQQsFoUyOib4eWfI\&amp;scisf=4\&amp;ct=citation\&amp;cd=19\&amp;hl=en}
    \item Modelops: Cloud-based lifecycle management for reliable and trusted ai (2019) \url{https://scholar.googleusercontent.com/scholar.bib?q=info:QmdGA7izEWUJ:scholar.google.com/\&amp;output=citation\&amp;scisdr=CgXSgnCxEMTkn4zmv3w:AAGBfm0AAAAAZBjgp30tUya8c9UNvXjmOFbxF9EIs_Jp\&amp;scisig=AAGBfm0AAAAAZBjgp-V2s1T1EEB2keqwdifh5pxGSD8g\&amp;scisf=4\&amp;ct=citation\&amp;cd=25\&amp;hl=en}
    \item The rise and evolution of agile software development (2018) \url{https://scholar.googleusercontent.com/scholar.bib?q=info:4X1OVYLmVCYJ:scholar.google.com/\&amp;output=citation\&amp;scisdr=CgXSgnCxEMTkn4zZZMc:AAGBfm0AAAAAZBjffMZ5GAp9IF_DRofs2ecgV6FcEgb0\&amp;scisig=AAGBfm0AAAAAZBjffJyuYLx7LKELwdBIKDV00O5Gf4L7\&amp;scisf=4\&amp;ct=citation\&amp;cd=1\&amp;hl=en}
    \item Using Continuous Integration to organize and monitor the annotation process of domain specific corpora (2014) \url{https://scholar.googleusercontent.com/scholar.bib?q=info:_8HKkIJfZVoJ:scholar.google.com/\&amp;output=citation\&amp;scisdr=CgXSgnCxEMTkn4zZZMc:AAGBfm0AAAAAZBjffMZ5GAp9IF_DRofs2ecgV6FcEgb0\&amp;scisig=AAGBfm0AAAAAZBjffP-7zl8KLXvvkaCP1Ls-_qKB5XfB\&amp;scisf=4\&amp;ct=citation\&amp;cd=6\&amp;hl=en}
    \item A hybrid model for agile practices using case based reasoning (2013) \url{https://scholar.googleusercontent.com/scholar.bib?q=info:tTi3jHnCKeAJ:scholar.google.com/\&amp;output=citation\&amp;scisdr=CgXSgnCxEMTkn4zZZMc:AAGBfm0AAAAAZBjffMZ5GAp9IF_DRofs2ecgV6FcEgb0\&amp;scisig=AAGBfm0AAAAAZBjffDycJDOWKaxmOYizMI4iKz3A8Te4\&amp;scisf=4\&amp;ct=citation\&amp;cd=10\&amp;hl=en}
    \item Knowledge engineering support for software requirements, architectures and components (2010) \url{https://scholar.googleusercontent.com/scholar.bib?q=info:UxhJUruMuWEJ:scholar.google.com/\&amp;output=citation\&amp;scisdr=CgXSgnCxEMTkn4zmv3w:AAGBfm0AAAAAZBjgp30tUya8c9UNvXjmOFbxF9EIs_Jp\&amp;scisig=AAGBfm0AAAAAZBjgp-SAk779Wsf1VOvkFSPxRNUkuyfV\&amp;scisf=4\&amp;ct=citation\&amp;cd=21\&amp;hl=en}
    \item Teaching agile software development: A case study (2010) \url{https://scholar.googleusercontent.com/scholar.bib?q=info:sPrlaSUoM40J:scholar.google.com/\&amp;output=citation\&amp;scisdr=CgXSgnCxEMTkn4zmv3w:AAGBfm0AAAAAZBjgp30tUya8c9UNvXjmOFbxF9EIs_Jp\&amp;scisig=AAGBfm0AAAAAZBjgp7UsLP-GTih4xbyyiRhqkvcRHHji\&amp;scisf=4\&amp;ct=citation\&amp;cd=31\&amp;hl=en}
    \item The role of project management in ineffective decision making within Agile software development projects (2009) \url{https://scholar.googleusercontent.com/scholar.bib?q=info:Z6ttxeKwxegJ:scholar.google.com/\&amp;output=citation\&amp;scisdr=CgXSgnCxEMTkn4zmv3w:AAGBfm0AAAAAZBjgp30tUya8c9UNvXjmOFbxF9EIs_Jp\&amp;scisig=AAGBfm0AAAAAZBjgp-t9HHO9agz4DF-PqpVTWTa95wcg\&amp;scisf=4\&amp;ct=citation\&amp;cd=23\&amp;hl=en}
    \item Empirical studies of agile software development: A systematic review (2008) \url{https://scholar.googleusercontent.com/scholar.bib?q=info:Tgk3kojMeu0J:scholar.google.com/\&amp;output=citation\&amp;scisdr=CgXSgnCxEMTkn4zmv3w:AAGBfm0AAAAAZBjgp30tUya8c9UNvXjmOFbxF9EIs_Jp\&amp;scisig=AAGBfm0AAAAAZBjgp0vTX4gbvAMT2wHccGuffIiCAwfe\&amp;scisf=4\&amp;ct=citation\&amp;cd=22\&amp;hl=en}
\end{itemize}

\noindent \textbf{IEEE Xplore}:
\begin{itemize}
    \item GPT2SP: A Transformer-Based Agile Story Point Estimation Approach (2023) \url{https://ieeexplore.ieee.org/stamp/stamp.jsp?arnumber=9732669}
    \item Neural Transfer Learning for Repairing Security Vulnerabilities in C Code (2023) \url{https://ieeexplore.ieee.org/stamp/stamp.jsp?arnumber=9699412}
    \item Toward Ambient Intelligence: Federated Edge Learning With Task-Oriented Sensing, Computation, and Communication Integration (2023) \url{https://ieeexplore.ieee.org/stamp/stamp.jsp?arnumber=9970330}
    \item Big Data Life Cycle in Shop-floor – Trends and Challenges (2023) \url{https://ieeexplore.ieee.org/abstract/document/10061223}
    \item A Review of Effort Estimation in Agile Software Development using Machine Learning Techniques (2022) \url{https://ieeexplore.ieee.org/stamp/stamp.jsp?arnumber=9985542}
    \item Machine Learning Regression Techniques for Test Case Prioritization in Continuous Integration Environment (2022) \url{https://ieeexplore.ieee.org/stamp/stamp.jsp?arnumber=9825820}
    \item Factors Affecting On-Time Delivery in Large-Scale Agile Software Development (2022) \url{https://ieeexplore.ieee.org/stamp/stamp.jsp?arnumber=9503331}
    \item MLOps: A Taxonomy and a Methodology (2022) \url{https://ieeexplore.ieee.org/stamp/stamp.jsp?arnumber=9792270}
    \item User Story Splitting in Agile Software Development using Machine Learning Approach (2022) \url{https://ieeexplore.ieee.org/stamp/stamp.jsp?arnumber=10053226}
    \item What Is an AI Engineer? An Empirical Analysis of Job Ads in The Netherlands (2022) \url{https://ieeexplore.ieee.org/stamp/stamp.jsp?arnumber=9796408}
    \item Challenges in Machine Learning Application Development: An Industrial Experience Report (2022) \url{https://ieeexplore.ieee.org/stamp/stamp.jsp?arnumber=9808693}
    \item Scaled Agile Framework Implementation in Organizations', its Shortcomings and an AI Based Solution to Track Team's Performance (2022) \url{https://ieeexplore.ieee.org/stamp/stamp.jsp?arnumber=9971968}
    \item Data Sovereignty for AI Pipelines: Lessons Learned from an Industrial Project at Mondragon Corporation (2022) \url{https://ieeexplore.ieee.org/stamp/stamp.jsp?arnumber=9796393}
    \item Jenkins Pipelines: A Novel Approach to Machine Learning Operations (MLOps) (2022) \url{https://ieeexplore.ieee.org/stamp/stamp.jsp?arnumber=9936252}
    \item Software Effort Estimation for Agile Software Development Using a Strategy Based on k-Nearest Neighbors Algorithm (2022) \url{https://ieeexplore.ieee.org/stamp/stamp.jsp?arnumber=9882947}
    \item Static Code Analysis Alarms Filtering Reloaded: A New Real-World Dataset and its ML-Based Utilization (2022) \url{https://ieeexplore.ieee.org/stamp/stamp.jsp?arnumber=9779763}
    \item Challenges Faced by Industries and Their Potential Solutions in Deploying Machine Learning Applications (2022) \url{https://ieeexplore.ieee.org/stamp/stamp.jsp?arnumber=9720900}
    \item Toward Integrating Intelligence and Programmability in Open Radio Access Networks: A Comprehensive Survey (2022) \url{https://ieeexplore.ieee.org/stamp/stamp.jsp?arnumber=9798822}
    \item Automated Risk Management Based Software Security Vulnerabilities Management (2022) \url{https://ieeexplore.ieee.org/stamp/stamp.jsp?arnumber=9802103}
    \item GELAB â€“ The Cutting Edge of Grammatical Evolution (2022) \url{https://ieeexplore.ieee.org/stamp/stamp.jsp?arnumber=9751757}
    \item Construction of Innovation and Entrepreneurship Information Sharing Platform Based on Multi-Dimensional Dynamic Innovation Model (2022) \url{https://ieeexplore.ieee.org/stamp/stamp.jsp?arnumber=10060627}
    \item Beyond 100 Ethical Concerns in the Development of Robot-to-Robot Cooperation (2022) \url{https://ieeexplore.ieee.org/stamp/stamp.jsp?arnumber=9903122}
    \item RLOps: Development Life-Cycle of Reinforcement Learning Aided Open RAN (2022) \url{https://ieeexplore.ieee.org/stamp/stamp.jsp?arnumber=9931127}
    \item Reinforcement Learning Reward Function for Test Case Prioritization in Continuous Integration (2022) \url{https://ieeexplore.ieee.org/stamp/stamp.jsp?arnumber=9756464}
    \item Deep Learning for B5G Open Radio Access Network: Evolution, Survey, Case Studies, and Challenges (2022) \url{https://ieeexplore.ieee.org/stamp/stamp.jsp?arnumber=9695955}
    \item Implementation of Agile Scrum Methodology in P4AI Application Development (2022) \url{https://ieeexplore.ieee.org/stamp/stamp.jsp?arnumber=10009692}
    \item Evaluation of Context-Aware Language Models and Experts for Effort Estimation of Software Maintenance Issues (2022) \url{https://ieeexplore.ieee.org/stamp/stamp.jsp?arnumber=9978209}
    \item Dynamic Pricing for Idle Resource in Public Clouds: Guarantee Revenue from Strategic Users (2022) \url{https://ieeexplore.ieee.org/stamp/stamp.jsp?arnumber=9798260}
    \item Designing and Developing a Data Science Programme in Bhutan (2022) \url{https://ieeexplore.ieee.org/stamp/stamp.jsp?arnumber=9962738}
    \item A Review of Effort Estimation in Agile Software Development using Machine Learning Techniques (2022) \url{https://ieeexplore.ieee.org/abstract/document/9985542/}
    \item Machine Learning Regression Techniques for Test Case Prioritization in Continuous Integration Environment (2022) \url{https://ieeexplore.ieee.org/abstract/document/9825820/}
    \item Challenges in Machine Learning Application Development: An Industrial Experience Report (2022) \url{https://ieeexplore.ieee.org/abstract/document/9808693/}
    \item Scaled Agile Framework Implementation in Organizations', its Shortcomings and an AI Based Solution to Track Team's Performance (2022) \url{https://ieeexplore.ieee.org/abstract/document/9971968/}
    \item Jenkins Pipelines: A Novel Approach to Machine Learning Operations (MLOps) (2022) \url{https://ieeexplore.ieee.org/abstract/document/9936252/}
    \item Software Effort Estimation for Agile Software Development Using a Strategy Based on k-Nearest Neighbors Algorithm (2022) \url{https://ieeexplore.ieee.org/abstract/document/9779763/}
    \item Artificial Intelligence based Risk Management Framework for Distributed Agile Software Development (2021) \url{https://ieeexplore.ieee.org/stamp/stamp.jsp?arnumber=9566000}
    \item Task Allocation in Distributed Agile Software Development using Machine Learning Approach (2021) \url{https://ieeexplore.ieee.org/stamp/stamp.jsp?arnumber=9688114}
    \item Advancing Design and Runtime Management of AI Applications with AI-SPRINT (Position Paper) (2021) \url{https://ieeexplore.ieee.org/stamp/stamp.jsp?arnumber=9529477}
    \item Framework for Machine Learning Based Task Allocation in DASD Environment (2021) \url{https://ieeexplore.ieee.org/stamp/stamp.jsp?arnumber=9514976}
    \item An Agile Software Development Life Cycle Model for Machine Learning Application Development (2021) \url{https://ieeexplore.ieee.org/stamp/stamp.jsp?arnumber=9664736}
    \item MLOps Challenges in Multi-Organization Setup: Experiences from Two Real-World Cases (2021) \url{https://ieeexplore.ieee.org/stamp/stamp.jsp?arnumber=9474388}
    \item TSAI - Test Selection using Artificial Intelligence for the Support of Continuous Integration (2021) \url{https://ieeexplore.ieee.org/stamp/stamp.jsp?arnumber=9700356}
    \item Who Needs MLOps: What Data Scientists Seek to Accomplish and How Can MLOps Help? (2021) \url{https://ieeexplore.ieee.org/stamp/stamp.jsp?arnumber=9474355}
    \item Weighted Reward for Reinforcement Learning based Test Case Prioritization in Continuous Integration Testing (2021) \url{https://ieeexplore.ieee.org/stamp/stamp.jsp?arnumber=9529787}
    \item A Network Intelligence Architecture for Efficient VNF Lifecycle Management (2021) \url{https://ieeexplore.ieee.org/stamp/stamp.jsp?arnumber=9163084}
    \item Quality-Aware DevOps Research: Where Do We Stand? (2021) \url{https://ieeexplore.ieee.org/stamp/stamp.jsp?arnumber=9373305}
    \item Technical Briefing: Hands-On Session on the Development of Trustworthy AI Software (2021) \url{https://ieeexplore.ieee.org/stamp/stamp.jsp?arnumber=9402396}
    \item Extending SOUP to ML Models When Designing Certified Medical Systems (2021) \url{https://ieeexplore.ieee.org/stamp/stamp.jsp?arnumber=9470897}
    \item The Promotion of Artificial Intelligence to the Development of the Sports Industry (2021) \url{https://ieeexplore.ieee.org/stamp/stamp.jsp?arnumber=9759961}
    \item Automated Counterfactual Generation in Financial Model Risk Management (2021) \url{https://ieeexplore.ieee.org/stamp/stamp.jsp?arnumber=9671561}
    \item DeepOrder: Deep Learning for Test Case Prioritization in Continuous Integration Testing (2021) \url{https://ieeexplore.ieee.org/stamp/stamp.jsp?arnumber=9609187}
    \item Robot Testing from a machine learning perspective (2021) \url{https://ieeexplore.ieee.org/stamp/stamp.jsp?arnumber=9698727}
    \item AgileML: A Machine Learning Project Development Pipeline Incorporating Active Consumer Engagement (2021) \url{https://ieeexplore.ieee.org/stamp/stamp.jsp?arnumber=9718470}
    \item Research on the Development of Library in the Era of Artificial Intelligence (2021) \url{https://ieeexplore.ieee.org/stamp/stamp.jsp?arnumber=9569594}
    \item Edge MLOps: An Automation Framework for AIoT Applications (2021) \url{https://ieeexplore.ieee.org/stamp/stamp.jsp?arnumber=9610376}
    \item An Edge Intelligent Framework for O-RAN based IoV Networks (2021) \url{https://ieeexplore.ieee.org/stamp/stamp.jsp?arnumber=9625234}
    \item Machine Learning-Assisted Analysis of Small Angle X-ray Scattering (2021) \url{https://ieeexplore.ieee.org/stamp/stamp.jsp?arnumber=9638297}
    \item MuDelta: Delta-Oriented Mutation Testing at Commit Time (2021) \url{https://ieeexplore.ieee.org/stamp/stamp.jsp?arnumber=9402071}
    \item Assessing the Risk of Software Development in Agile Methodologies Using Simulation (2021) \url{https://ieeexplore.ieee.org/stamp/stamp.jsp?arnumber=9548910}
    \item Multi-agent cross prompt mechanism based on Actor-Critic gradient strategy (2021) \url{https://ieeexplore.ieee.org/stamp/stamp.jsp?arnumber=9515871}
    \item ACE: An ATAK Plugin for Enhanced Acoustic Situational Awareness at the Edge (2021) \url{https://ieeexplore.ieee.org/stamp/stamp.jsp?arnumber=9653127}
    \item FPGA-Based Real-Time Data Acquisition for Ultrafast X-Ray Computed Tomography (2021) \url{https://ieeexplore.ieee.org/stamp/stamp.jsp?arnumber=9591646}
    \item Design of Intelligent Auxiliary Input System for Distribution Equipment Defects Based on HMM and Word2vec (2021) \url{https://ieeexplore.ieee.org/stamp/stamp.jsp?arnumber=9736056}
    \item Big Data Resource Management \& Networks: Taxonomy, Survey, and Future Directions (2021) \url{https://ieeexplore.ieee.org/stamp/stamp.jsp?arnumber=9478917}
    \item Comprehensive Instructional Video Analysis: The COIN Dataset and Performance Evaluation (2021) \url{https://ieeexplore.ieee.org/stamp/stamp.jsp?arnumber=9037088}
    \item Artificial Intelligence based Risk Management Framework for Distributed Agile Software Development (2021) \url{https://ieeexplore.ieee.org/abstract/document/9566000/}
    \item Task allocation in distributed agile software development using machine learning approach (2021) \url{https://ieeexplore.ieee.org/abstract/document/9688114/}
    \item Advancing design and runtime management of AI applications with AI-SPRINT (position paper) (2021) \url{https://ieeexplore.ieee.org/abstract/document/9529477/}
    \item Framework for Machine Learning Based Task Allocation in DASD Environment (2021) \url{https://ieeexplore.ieee.org/abstract/document/9514976/}
    \item An Agile Software Development Life Cycle Model for Machine Learning Application Development (2021) \url{https://ieeexplore.ieee.org/abstract/document/9664736/}
    \item MLOps challenges in multi-organization setup: Experiences from two real-world cases (2021) \url{https://ieeexplore.ieee.org/abstract/document/9474388/}
    \item TSAI-Test Selection using Artificial Intelligence for the Support of Continuous Integration (2021) \url{https://ieeexplore.ieee.org/abstract/document/9700356/}
    \item Who needs MLOps: What data scientists seek to accomplish and how can MLOps help? (2021) \url{https://ieeexplore.ieee.org/abstract/document/9474355/}
    \item Weighted Reward for Reinforcement Learning based Test Case Prioritization in Continuous Integration Testing (2021) \url{https://ieeexplore.ieee.org/abstract/document/9529787/}
    \item Technical briefing: Hands-on session on the development of trustworthy AI software (2021) \url{https://ieeexplore.ieee.org/abstract/document/9402396/}
    \item Extending SOUP to ML models when designing certified medical systems (2021) \url{https://ieeexplore.ieee.org/abstract/document/9470897/}
    \item The Promotion of Artificial Intelligence to the Development of the Sports Industry (2021) \url{https://ieeexplore.ieee.org/abstract/document/9759961/}
    \item Automated Counterfactual Generation in Financial Model Risk Management (2021) \url{https://ieeexplore.ieee.org/abstract/document/9671561/}
    \item DeepOrder: Deep learning for test case prioritization in continuous integration testing (2021) \url{https://ieeexplore.ieee.org/abstract/document/9609187/}
    \item Neural Network Classification for Improving Continuous Regression Testing (2020) \url{https://ieeexplore.ieee.org/stamp/stamp.jsp?arnumber=9176804}
    \item Machine Learning Application in LAPIS Agile Software Development Process (2020) \url{https://ieeexplore.ieee.org/stamp/stamp.jsp?arnumber=9247069}
    \item DevOps for AI â€“ Challenges in Development of AI-enabled Applications (2020) \url{https://ieeexplore.ieee.org/stamp/stamp.jsp?arnumber=9238323}
    \item Machine Learning models to predict Agile Methodology adoption (2020) \url{https://ieeexplore.ieee.org/stamp/stamp.jsp?arnumber=9222987}
    \item Identifying and Generating Missing Tests using Machine Learning on Execution Traces (2020) \url{https://ieeexplore.ieee.org/stamp/stamp.jsp?arnumber=9176745}
    \item Towards a Secure Proxy-based Architecture for Collaborative AI Engineering (2020) \url{https://ieeexplore.ieee.org/stamp/stamp.jsp?arnumber=9355887}
    \item Learning-to-Rank vs Ranking-to-Learn: Strategies for Regression Testing in Continuous Integration (2020) \url{https://ieeexplore.ieee.org/stamp/stamp.jsp?arnumber=9283979}
    \item Quality of Service Measurement and Prediction through AI Technology (2020) \url{https://ieeexplore.ieee.org/stamp/stamp.jsp?arnumber=9302008}
    \item AD4ML: Axiomatic Design to Specify Machine Learning Solutions for Manufacturing (2020) \url{https://ieeexplore.ieee.org/stamp/stamp.jsp?arnumber=9191629}
    \item Using Machine Learning to Identify Code Fragments for Manual Review (2020) \url{https://ieeexplore.ieee.org/stamp/stamp.jsp?arnumber=9226313}
    \item Selective Regression Testing based on Big Data: Comparing Feature Extraction Techniques (2020) \url{https://ieeexplore.ieee.org/stamp/stamp.jsp?arnumber=9155577}
    \item Comparison of DotKernel and Symfony as PHP frameworks (2020) \url{https://ieeexplore.ieee.org/stamp/stamp.jsp?arnumber=9223157}
    \item Controlled time series generation for automotive software-in-the-loop testing using GANs (2020) \url{https://ieeexplore.ieee.org/stamp/stamp.jsp?arnumber=9176782}
    \item Accuracy Measurement of Deep Neural Network Accelerator via Metamorphic Testing (2020) \url{https://ieeexplore.ieee.org/stamp/stamp.jsp?arnumber=9176788}
    \item MIMO Cross-Layer Secure Communication Algorithm for Cyber Physical Systems Based on Interference Strategies (2020) \url{https://ieeexplore.ieee.org/stamp/stamp.jsp?arnumber=9296797}
    \item Towards a Competence Profile for Automotive Software Engineering (2020) \url{https://ieeexplore.ieee.org/stamp/stamp.jsp?arnumber=9206205}
    \item Characterizing the Occurrence of Dockerfile Smells in Open-Source Software: An Empirical Study (2020) \url{https://ieeexplore.ieee.org/stamp/stamp.jsp?arnumber=8998208}
    \item SDN Enhanced Resource Orchestration of Containerized Edge Applications for Industrial IoT (2020) \url{https://ieeexplore.ieee.org/stamp/stamp.jsp?arnumber=9296769}
    \item Use of Different Channels for User Awareness and Education Related to Fraud and Phishing in a Banking Institution (2020) \url{https://ieeexplore.ieee.org/stamp/stamp.jsp?arnumber=9379220}
    \item Sharing at Scale: An Open-Source-Software-based License Compliance Ecosystem (2020) \url{https://ieeexplore.ieee.org/stamp/stamp.jsp?arnumber=9276613}
    \item A network intelligence architecture for efficient vnf lifecycle management (2020) \url{https://ieeexplore.ieee.org/abstract/document/9163084/}
    \item Identifying and generating missing tests using machine learning on execution traces (2020) \url{https://ieeexplore.ieee.org/abstract/document/9176745/}
    \item Neural Network Classification for Improving Continuous Regression Testing (2020) \url{https://ieeexplore.ieee.org/abstract/document/9176804/}
    \item Machine learning models to predict agile methodology adoption (2020) \url{https://ieeexplore.ieee.org/abstract/document/9222987/}
    \item DevOps for AI--Challenges in Development of AI-enabled Applications (2020) \url{https://ieeexplore.ieee.org/abstract/document/9238323/}
    \item Quality of service measurement and prediction through AI technology (2020) \url{https://ieeexplore.ieee.org/abstract/document/9302008/}
    \item Towards a secure proxy-based architecture for collaborative AI engineering (2020) \url{https://ieeexplore.ieee.org/abstract/document/9355887/}
    \item Modelops: Cloud-based lifecycle management for reliable and trusted ai (2019) \url{https://ieeexplore.ieee.org/abstract/document/8790192/}
    \item Predictive Test Selection (2019) \url{https://ieeexplore.ieee.org/stamp/stamp.jsp?arnumber=8804462}
    \item Automated Trainability Evaluation for Smart Software Functions (2019) \url{https://ieeexplore.ieee.org/stamp/stamp.jsp?arnumber=8952173}
    \item ModelOps: Cloud-Based Lifecycle Management for Reliable and Trusted AI (2019) \url{https://ieeexplore.ieee.org/stamp/stamp.jsp?arnumber=8790192}
    \item A Study on the Interplay between Pull Request Review and Continuous Integration Builds (2019) \url{https://ieeexplore.ieee.org/stamp/stamp.jsp?arnumber=8667996}
    \item Research on Construction of Cloud Computing Platform for Railway Enterprises (2019) \url{https://ieeexplore.ieee.org/stamp/stamp.jsp?arnumber=8950904}
    \item Remote Receiver Control in MPTCP Networks in Uncertain Operating Conditions (2019) \url{https://ieeexplore.ieee.org/stamp/stamp.jsp?arnumber=8864677}
    \item Predictive test selection (2019) \url{https://ieeexplore.ieee.org/abstract/document/8804462/}
    \item Automated trainability evaluation for smart software functions (2019) \url{https://ieeexplore.ieee.org/abstract/document/8952173/}
    \item Behave Nicely! Automatic Generation of Code for Behaviour Driven Development Test Suites (2019) \url{https://ieeexplore.ieee.org/document/8930836}
    \item Poster: ACONA: Active Online Model Adaptation for Predicting Continuous Integration Build Failures (2018) \url{https://ieeexplore.ieee.org/stamp/stamp.jsp?arnumber=8449581}
    \item A Survey of Software Quality for Machine Learning Applications (2018) \url{https://ieeexplore.ieee.org/stamp/stamp.jsp?arnumber=8411764}
    \item Incumbent Firm Capacity Building in Analytics: Strategy, Structure, and Innovation Management Perspectives (2018) \url{https://ieeexplore.ieee.org/stamp/stamp.jsp?arnumber=8481749}
    \item A survey of software quality for machine learning applications (2018) \url{https://ieeexplore.ieee.org/abstract/document/9191629/}
    \item What do we (really) know about test-driven development? (2018) \url{https://ieeexplore.ieee.org/document/8405634}
    \item d(mu)Reg: A Path-Aware Mutation Analysis Guided Approach to Regression Testing (2017) \url{https://ieeexplore.ieee.org/stamp/stamp.jsp?arnumber=7962333}
    \item Built to Last or Built Too Fast? Evaluating Prediction Models for Build Times (2017) \url{https://ieeexplore.ieee.org/stamp/stamp.jsp?arnumber=7962403}
    \item End-to-end service data analysis: Efficiencies achieved across the enterprise (2017) \url{https://ieeexplore.ieee.org/stamp/stamp.jsp?arnumber=7877285}
    \item A roadmap to the programmable world: software challenges in the IoT era (2017) \url{https://ieeexplore.ieee.org/document/7819416}
    \item Using ontology to enhance requirement engineering in agile software process (2016) \url{https://ieeexplore.ieee.org/stamp/stamp.jsp?arnumber=7916218}
    \item Using Continuous Integration to organize and monitor the annotation process of domain specific corpora (2014) \url{https://ieeexplore.ieee.org/abstract/document/6841958/}
    \item Using Continuous Integration to organize and monitor the annotation process of domain specific corpora (2014) \url{https://ieeexplore.ieee.org/stamp/stamp.jsp?arnumber=6841958}
    \item Incorporating artificial intelligence technique into DSDM (2014) \url{https://ieeexplore.ieee.org/stamp/stamp.jsp?arnumber=7053838}
    \item E-health decision support system for differential diagnosis (2014) \url{https://ieeexplore.ieee.org/stamp/stamp.jsp?arnumber=6857834}
    \item A hybrid model for agile practices using case based reasoning (2013) \url{https://ieeexplore.ieee.org/abstract/document/6615431/}
    \item A hybrid model for agile practices using case based reasoning (2013) \url{https://ieeexplore.ieee.org/stamp/stamp.jsp?arnumber=6615431}
    \item Local versus global lessons for defect prediction and effort estimation (2012) \url{https://ieeexplore.ieee.org/document/6363444}
\end{itemize}

\noindent \textbf{ScienceDirect}:
\begin{itemize}
    \item Chapter 21 - Model management and ModelOps: managing an artificial intelligence-driven enterpriseâ˜†â˜†Note: TIBCO and Omni-HealthData are trademarks or registered trademarks of TIBCO Software Inc. and/or its subsidiaries in the United States and/or other countries. (2023) \url{https://www.sciencedirect.com/science/article/pii/B9780323952743000269}
    \item Chapter 13 - Emerging clouds (2023) \url{https://www.sciencedirect.com/science/article/pii/B9780323852777000208}
    \item A comprehensive evaluation framework for deep model robustness (2023) \url{https://www.sciencedirect.com/science/article/pii/S0031320323000092}
    \item Continuous deployment in software-intensive system-of-systems (2023) \url{https://www.sciencedirect.com/science/article/pii/S095058492300054X}
    \item Towards cost-effective and robust AI microservice deployment in edge computing environments (2023) \url{https://www.sciencedirect.com/science/article/pii/S0167739X22003314}
    \item Introduction to special issue on Agile UX: challenges, successes and barriers to improvement (2023) \url{https://www.sciencedirect.com/science/article/pii/S0950584923000472}
    \item Artificial intelligence for industry 4.0: Systematic review of applications, challenges, and opportunities (2023) \url{https://www.sciencedirect.com/science/article/pii/S0957417422024757}
    \item Knowledge diffusion trajectories of agile software development research: A main path analysis (2023) \url{https://www.sciencedirect.com/science/article/pii/S0950584922002403}
    \item 5G/5G+ network management employing AI-based continuous deployment (2023) \url{https://www.sciencedirect.com/science/article/pii/S1568494623000029}
    \item Chapter 1 - The essence of intelligence (2023) \url{https://www.sciencedirect.com/science/article/pii/B9780323995627000012}
    \item Chapter 9 - Reflection on human-machine hybrid intelligence (2023) \url{https://www.sciencedirect.com/science/article/pii/B9780323995627000097}
    \item Orfeon: An AIOps framework for the goal-driven operationalization of distributed analytical pipelines (2023) \url{https://www.sciencedirect.com/science/article/pii/S0167739X22003223}
    \item A conceptual model supporting decision-making for test automation in Agile-based Software Development (2023) \url{https://www.sciencedirect.com/science/article/pii/S0169023X22001021}
    \item Short-term synaptic plasticity in emerging devices for neuromorphic computing (2023) \url{https://www.sciencedirect.com/science/article/pii/S2589004223003929}
    \item Steering resilience in nursing practice: Examining the impact of digital innovations and enhanced emotional training on nurse competencies (2023) \url{https://www.sciencedirect.com/science/article/pii/S0166497222000967}
    \item Deep learning for detecting macroplastic litter in water bodies: A review (2023) \url{https://www.sciencedirect.com/science/article/pii/S0043135423000672}
    \item Transparency and Explainability of AI Systems: From Ethical Guidelines to Requirements (2023) \url{https://www.sciencedirect.com/science/article/pii/S0950584923000514}
    \item Worker and workplace Artificial Intelligence (AI) coexistence: Emerging themes and research agenda (2023) \url{https://www.sciencedirect.com/science/article/pii/S0166497223000585}
    \item Data collection, wrangling, and pre-processing for AI assurance (2023) \url{https://www.sciencedirect.com/science/article/pii/B9780323919197000226}
    \item Role of artificial intelligence in environmental sustainability (2023) \url{https://www.sciencedirect.com/science/article/pii/B9780323997140000091}
    \item Framework for automatic detection of anomalies in DevOps (2023) \url{https://www.sciencedirect.com/science/article/pii/S1319157823000393}
    \item The pipeline for the continuous development of artificial intelligence modelsâ€”Current state of research and practice (2023) \url{https://www.sciencedirect.com/science/article/pii/S0164121223000109}
    \item 5G/5G+ network management employing AI-based continuous deployment (2023) \url{https://www.sciencedirect.com/science/article/pii/S1568494623000029}
    \item FAT Forensics: A Python toolbox for algorithmic fairness, accountability and transparency (2022) \url{https://www.sciencedirect.com/science/article/pii/S2665963822000951}
    \item Guideline for software life cycle in health informatics (2022) \url{https://www.sciencedirect.com/science/article/pii/S2589004222018065}
    \item Multilayer perceptron neural network model development for mechanical ventilator parameters prediction by real time system learning (2022) \url{https://www.sciencedirect.com/science/article/pii/S1746809421007679}
    \item Chapter One - Exploring the edge AI space: Industry use cases (2022) \url{https://www.sciencedirect.com/science/article/pii/S0065245822000304}
    \item A systematic literature review on software defect prediction using artificial intelligence: Datasets, Data Validation Methods, Approaches, and Tools (2022) \url{https://www.sciencedirect.com/science/article/pii/S0952197622000616}
    \item Kafka-ML: Connecting the data stream with ML/AI frameworks (2022) \url{https://www.sciencedirect.com/science/article/pii/S0167739X21002995}
    \item AIDOaRt: AI-augmented Automation for DevOps, a model-based framework for continuous development in Cyberâ€“Physical Systems (2022) \url{https://www.sciencedirect.com/science/article/pii/S0141933122002022}
    \item Artificial intelligence in tourism and hospitality: Bibliometric analysis and research agenda (2022) \url{https://www.sciencedirect.com/science/article/pii/S0278431922001797}
    \item histolab: A Python library for reproducible Digital Pathology preprocessing with automated testing (2022) \url{https://www.sciencedirect.com/science/article/pii/S2352711022001558}
    \item Decision Support Collaborative Platform for e-Health Integration in Smart Communities Context (2022) \url{https://www.sciencedirect.com/science/article/pii/S1877050922020026}
    \item Viral outbreaks detection and surveillance using wastewater-based epidemiology, viral air sampling, and machine learning techniques: A comprehensive review and outlook (2022) \url{https://www.sciencedirect.com/science/article/pii/S0048969721049093}
    \item Introduction to the special issue on managing software processes using soft computing techniques (2022) \url{https://www.sciencedirect.com/science/article/pii/S0950584922001665}
    \item A Trusted Platform Module-based, Pre-emptive and Dynamic Asset Discovery Tool (2022) \url{https://www.sciencedirect.com/science/article/pii/S2214212622001958}
    \item The moon, the ghetto and artificial intelligence: Reducing systemic racism in computational algorithms (2022) \url{https://www.sciencedirect.com/science/article/pii/S0740624X21000812}
    \item Test automation maturity improves product qualityâ€”Quantitative study of open source projects using continuous integration (2022) \url{https://www.sciencedirect.com/science/article/pii/S0164121222000280}
    \item An artificial intelligence life cycle: From conception to production (2022) \url{https://www.sciencedirect.com/science/article/pii/S2666389922000745}
    \item Smart manufacturing powered by recent technological advancements: A review (2022) \url{https://www.sciencedirect.com/science/article/pii/S0278612522001042}
    \item Guideline for Deployment of Machine Learning Models for Predictive Quality in Production (2022) \url{https://www.sciencedirect.com/science/article/pii/S2212827122003523}
    \item AI-driven laboratory workflows enable operation in the age of social distancing (2022) \url{https://www.sciencedirect.com/science/article/pii/S2472630321000236}
    \item Functional precision cancer medicine: drug sensitivity screening enabled by cell culture models (2022) \url{https://www.sciencedirect.com/science/article/pii/S016561472200195X}
    \item Artificial Intelligence and COVID-19: A Systematic umbrella review and roads ahead (2022) \url{https://www.sciencedirect.com/science/article/pii/S1319157821001774}
    \item Using artificial intelligence to support the drawing of piping and instrumentation diagrams using DEXPI standard (2022) \url{https://www.sciencedirect.com/science/article/pii/S2772508122000291}
    \item Quality measurement in agile and rapid software development: A systematic mapping (2022) \url{https://www.sciencedirect.com/science/article/pii/S0164121221002661}
    \item A deep learning-based automated framework for functional User Interface testing (2022) \url{https://www.sciencedirect.com/science/article/pii/S0950584922001070}
    \item Tailoring the Scrum framework for software development: Literature mapping and feature-based support (2022) \url{https://www.sciencedirect.com/science/article/pii/S0950584921002457}
    \item Smart manufacturing powered by recent technological advancements: A review (2022) \url{https://www.sciencedirect.com/science/article/pii/S0278612522001042}
    \item A workflow for the sustainable development of closure models for bubbly flows (2021) \url{https://www.sciencedirect.com/science/article/pii/S0009250921003729}
    \item 9 - Covid-19 accelerating the dynamics of Artificial Intelligence disruption (2021) \url{https://www.sciencedirect.com/science/article/pii/B978032389777800004X}
    \item BeeToxAI: An artificial intelligence-based web app to assess acute toxicity of chemicals to honey bees (2021) \url{https://www.sciencedirect.com/science/article/pii/S2667318521000131}
    \item Business Intelligence Capabilities and Firm Performance: A Study in China (2021) \url{https://www.sciencedirect.com/science/article/pii/S0268401220314316}
    \item Integration of agile practices in the product development process of intelligent technical systems (2021) \url{https://www.sciencedirect.com/science/article/pii/S2212827121005679}
    \item Big Data analytics in Agile software development: A systematic mapping study (2021) \url{https://www.sciencedirect.com/science/article/pii/S0950584920301981}
    \item Fostering reproducibility, reusability, and technology transfer in health informatics (2021) \url{https://www.sciencedirect.com/science/article/pii/S2589004221007719}
    \item Influence of new-age technologies on marketing: A research agenda (2021) \url{https://www.sciencedirect.com/science/article/pii/S0148296320300151}
    \item Ready, Steady, Go AI: A practical tutorial on fundamentals of artificial intelligence and its applications in phenomics image analysis (2021) \url{https://www.sciencedirect.com/science/article/pii/S2666389921001719}
    \item Development of an AI-based expert system for the part- and process-specific marking of materials (2021) \url{https://www.sciencedirect.com/science/article/pii/S2212827121005503}
    \item Application of machine learning and artificial intelligence in oil and gas industry (2021) \url{https://www.sciencedirect.com/science/article/pii/S2096249521000429}
    \item 7 - Project management for a machine learning project (2021) \url{https://www.sciencedirect.com/science/article/pii/B978012819742400007X}
    \item Analysis on severe fever with thrombocytopenia syndrome bunyavirus infection combined with atrial fibrillation under digital model detection (2021) \url{https://www.sciencedirect.com/science/article/pii/S2211379721005349}
    \item Towards a continuous certification of safety-critical avionics software (2021) \url{https://www.sciencedirect.com/science/article/pii/S0166361520306163}
    \item Large scale quality transformation in hybrid development organizations â€“ A case study (2021) \url{https://www.sciencedirect.com/science/article/pii/S0164121220302284}
    \item Adoption and Effects of Software Engineering Best Practices in Machine Learning (2020) \url{https://doi.org/10.1145/3382494.3410681}
    \item Using AI to Facilitate Technology Management â€“ Designing an Automated Technology Radar (2020) \url{https://www.sciencedirect.com/science/article/pii/S221282712030723X}
    \item Utilization of a reinforcement learning algorithm for the accurate alignment of a robotic arm in a complete soft fabric shoe tongues automation process (2020) \url{https://www.sciencedirect.com/science/article/pii/S0278612520301114}
    \item Modeling continuous security: A conceptual model for automated DevSecOps using open-source software over cloud (ADOC) (2020) \url{https://www.sciencedirect.com/science/article/pii/S0167404820302406}
    \item An extensive study of class-level and method-level test case selection for continuous integration (2020) \url{https://www.sciencedirect.com/science/article/pii/S0164121220300923}
    \item Social Responsibility, Human Centred Systems and Engineering Ethics:: A New Manifesto for Systems Engineering Design Praxis (2020) \url{https://www.sciencedirect.com/science/article/pii/S2405896320327695}
    \item The Role of Process Engineering in the Digital Transformation (2020) \url{https://www.sciencedirect.com/science/article/pii/B978012823377150121X}
    \item An agile co-creation process for digital servitization: A micro-service innovation approach (2020) \url{https://www.sciencedirect.com/science/article/pii/S0148296320300175}
    \item Opportunity for renewal or disruptive force? How artificial intelligence alters democratic politics (2020) \url{https://www.sciencedirect.com/science/article/pii/S0740624X1930245X}
    \item Intelligent software engineering in the context of agile software development: A systematic literature review (2020) \url{https://www.sciencedirect.com/science/article/pii/S0950584919302587}
    \item Chapter 9 - Software engineering (2020) \url{https://www.sciencedirect.com/science/article/pii/B9780081026946000097}
    \item Test Case Prioritization in Continuous Integration environments: A systematic mapping study (2020) \url{https://www.sciencedirect.com/science/article/pii/S0950584920300185}
    \item DevOps in Practice for Education Management Information System at ECNU (2020) \url{https://www.sciencedirect.com/science/article/pii/S1877050920320482}
    \item 10 - The application of artificial intelligence in software engineering: a review challenging conventional wisdom (2020) \url{https://www.sciencedirect.com/science/article/pii/B9780128183663000101}
    \item Chapter 17 - AI-driven process change (2019) \url{https://www.sciencedirect.com/science/article/pii/B9780128158470000170}
    \item Proposal of a Visual Environment to Support Scrum (2019) \url{https://www.sciencedirect.com/science/article/pii/S1877050919322586}
    \item Usability in agile software development: A tertiary study (2019) \url{https://www.sciencedirect.com/science/article/pii/S0920548918302587}
    \item The role of Sprint planning and feedback in game development projects: Implications for game quality (2019) \url{https://www.sciencedirect.com/science/article/pii/S0164121219300974}
    \item Advances in using agile and lean processes for software development (2019) \url{https://www.sciencedirect.com/science/article/abs/pii/S0065245818300299?via%3Dihub}
    \item An intelligent system on knowledge generation and communication about flooding (2018) \url{https://www.sciencedirect.com/science/article/pii/S1364815217308368}
    \item Agile development in the cloud computing environment: A systematic review (2018) \url{https://www.sciencedirect.com/science/article/pii/S0950584918301319}
    \item Requirements engineering: A systematic mapping study in agile software development (2018) \url{https://www.sciencedirect.com/science/article/pii/S0164121218300141}
    \item Adapting agile practices in university contexts (2018) \url{https://www.sciencedirect.com/science/article/pii/S0164121218301419}
    \item Developing and using checklists to improve software effort estimation: A multi-case study (2018) \url{https://www.sciencedirect.com/science/article/pii/S0164121218302073}
    \item Agile government: Systematic literature review and future research (2018) \url{https://www.sciencedirect.com/science/article/pii/S0740624X18302107}
    \item Chapter 4 - Making the Case for Artificial Intelligence at Government: Guidelines to Transforming Federal Software Systems (2018) \url{https://www.sciencedirect.com/science/article/pii/B9780128124437000041}
    \item AI and Big Data: A blueprint for a human rights, social and ethical impact assessment (2018) \url{https://www.sciencedirect.com/science/article/pii/S0267364918302012}
    \item Problems, causes and solutions when adopting continuous deliveryâ€”A systematic literature review (2017) \url{https://www.sciencedirect.com/science/article/pii/S0950584916302324}
    \item The RIGHT model for Continuous Experimentation (2017) \url{https://www.sciencedirect.com/science/article/pii/S0164121216300024}
    \item Continuous deployment of software intensive products and services: A systematic mapping study (2017) \url{https://www.sciencedirect.com/science/article/abs/pii/S0164121215002812}
    \item An exploratory study in communication in Agile Global Software Development (2016) \url{https://www.sciencedirect.com/science/article/abs/pii/S0920548916300381}
    \item Progressive Outcomes: A framework for maturing in agile software development (2015) \url{https://www.sciencedirect.com/science/article/pii/S0164121214002908}
    \item Agile Principles and Achievement of Success in Software Development: A Quantitative Study in Brazilian Organizations (2014) \url{https://www.sciencedirect.com/science/article/pii/S2212017314002485}
    \item Processes versus people: How should agile software development maturity be defined? (2014) \url{https://www.sciencedirect.com/science/article/pii/S0164121214001587}
    \item Modeling the leadership â€“ project performance relation: radial basis function, Gaussian and Kriging methods as alternatives to linear regression (2013) \url{https://www.sciencedirect.com/science/article/pii/S0957417412008688}
    \item Applying Bayesian networks to performance forecast of innovation projects: A case study of transformational leadership influence in organizations oriented by projects (2012) \url{https://www.sciencedirect.com/science/article/pii/S0957417411015764}
    \item A survey study of critical success factors in agile software projects (2008) \url{https://www.sciencedirect.com/science/article/pii/S0164121207002208}
    \item Design issues for knowledge artifacts (2008) \url{https://www.sciencedirect.com/science/article/pii/S0950705108000944}
    \item Analytical design for linear continuous integrator/time delay systems (1999) \url{https://www.sciencedirect.com/science/article/pii/S1474667017565064}
    \item Discrete and continuous integrable systems possessing the same non-dynamical r-matrix (1997) \url{https://www.sciencedirect.com/science/article/pii/S0375960197005379}
\end{itemize}

\noindent \textbf{Springer}:
\begin{itemize}
    \item Machine Learning Application in LAPIS Agile Software Development Process (2022) \url{https://link.springer.com/chapter/10.1007/978-3-030-80119-9_77}
    \item An industrial case study of classifier ensembles for locating software defects (2011) \url{https://link.springer.com/article/10.1007/s11219-010-9128-1}
\end{itemize}

\noindent \textbf{Wiley}: Systematic review of success factors and barriers for software process improvement in global software development (2016) \url{https://ietresearch.onlinelibrary.wiley.com/doi/10.1049/iet-sen.2015.0038}

%%%%%%%%%%%%%%%%%%%%%%%%%%%%%%%%%%%%%%%%%%%